\documentclass[%
 reprint,
superscriptaddress,
 amsmath,amssymb,longbibliography,
 aps,
prx,
]{revtex4-2}
\pdfoutput=1
\usepackage{hyperref}
\usepackage{graphicx}
\usepackage{amsmath}
\usepackage{color}
\usepackage{physics}
\usepackage{dsfont}
\usepackage{comment}
\usepackage{physics}
\usepackage{booktabs}

\newif\ifarXiv
\arXivtrue 

\definecolor{limegreen}{rgb}{0.2, 0.8, 0.2}
\definecolor{orange}{rgb}{1.0, 0.5, 0.0}

\definecolor{emerald}{rgb}{0.31, 0.78, 0.47}
\definecolor{blue(ncs)}{rgb}{0.0, 0.53, 0.74}

\hypersetup{
     colorlinks=true,
     linkcolor=magenta,
     filecolor=blue,
     citecolor=emerald,      
     urlcolor =blue(ncs),
}

\newcommand{\kv}{{\bf k}}
\newcommand{\qv}{{\bf q}}

\newcommand{\Gv}{{\bf G}}
\newcommand{\tv}{{\bf t}_{12}}
\begin{document}
\title{Minimal Models for Altermagnetism}

\author{Mercè Roig}
\affiliation{Niels Bohr Institute, University of Copenhagen, DK-2100 Copenhagen, Denmark}

\author{Andreas Kreisel}
\affiliation{Niels Bohr Institute, University of Copenhagen, DK-2100 Copenhagen, Denmark} 

\author{Yue Yu}
\affiliation{Department of Physics, University of Wisconsin–Milwaukee, Milwaukee, Wisconsin 53201, USA} 

\author{Brian M. Andersen}
\affiliation{Niels Bohr Institute, University of Copenhagen, DK-2100 Copenhagen, Denmark} 

\author{Daniel F. Agterberg}
\affiliation{Department of Physics, University of Wisconsin–Milwaukee, Milwaukee, Wisconsin 53201, USA} 


\vskip 1cm

\begin{abstract}
Altermagnets feature vanishing net magnetization, like antiferromagnets, but exhibit time-reversal symmetry breaking and momentum-dependent spin-split band structures. Motivated by the fact that all proposed altermagnets have paramagnetic states with multiple magnetic ions in the unit cell,  we develop a class of realistic minimal models for altermagnetism through a comparative analysis of the magnetic atom Wyckoff site symmetry and the space group symmetry. Specifically, we develop electronic models for all centrosymmetric space groups with  magnetic atoms occupying inversion symmetric Wyckoff positions with multiplicity two.  These forty models include monoclinic, orthorhombic, tetragonal, rhombohedral, hexagonal, and cubic materials and describe $d$-wave, $g$-wave, and $i$-wave altermagnetism. We further define and examine an altermagnetic susceptibility and mean field instabilities within a Hubbard model to  reveal that these models have altermagnetic ground states.  We shed insight on why most altermagnets form in non-symmorphic space groups. We also provide the symmetry-required form of the spin-orbit coupling and show it yields a Berry curvature that is {\it linear} in this  coupling for all forty models. We apply our models to representative cases of RuO$_2$, MnF$_2$, FeSb$_2$, $\kappa$-Cl, CrSb, and MnTe.
\end{abstract}

\maketitle

\section{Introduction}
Altermagnetism has been recently recognized as a new class of magnetic order~\cite{Hayami2019,Ahn2019,Smejkal2020Jun,Yuan2020Jul,Mazin2021,Smejkal2022Sep,Smejkal2022Dec,Bhowal2022Dec}. 
This exceptional state shares common features with both conventional ferromagnets and antiferromagnets. In particular, altermagnets exhibit energy splitting between spin states, similar to ferromagnets, while still featuring vanishing net magnetization, akin to antiferromagnets. 
Nevertheless, the opposite spin sublattices in an altermagnet are not related by translation or inversion, but instead they are connected by a crystal rotation symmetry. 
A large number of materials have been proposed to host this collinear-compensated magnetic order~\cite{Smejkal2022Sep,Smejkal2022Dec}, including the rutile metals RuO$_2$~\cite{Berlijn2017Feb,Ahn2019,Smejkal2020Jun} and MnF$_2$~\cite{Yuan2020Jul,Bhowal2022Dec}, FeSb$_2$~\cite{Mazin2021}, $\kappa$-Cl~\cite{Naka2019,Naka2020}, MnTe \cite{Lee2024} and CrSb~\cite{Reimers2024}. In Ref.~\onlinecite{Guo2023Mar}, a search through the MAGNDATA database of magnetic materials yields 62 altermagnetic candidate materials.

The unique electronic structure of altermagnets exhibiting a spin splitting in reciprocal space makes them candidates for spintronics applications~\cite{Yuan2020Jul,Gonzalez-Hernandez2021Mar,Smejkal2022Feb,Shao2021Dec}.
Specifically, the vanishing net magnetization consequently leads to insensitivity to external magnetic field perturbations, and allows for applications without requiring relativistic spin-orbit coupling (SOC).
Another consequence predicted for this time-reversal symmetry breaking phase includes the anomalous Hall effect~\cite{Smejkal2020Jun,Smejkal2022Jun}, previously associated mainly with ferromagnetism.
In the case of MnTe or RuO$_2$, for example, recent experimental reports are consistent with the expected response of altermagnets, including the crystal Hall effect~\cite{Feng2022Nov,Betancourt2023}, spin currents~\cite{Bose2022May}, spin-splitting torque phenomena~\cite{Bai2022May,Karube2022Sep}, and broken Kramer's degeneracy in the band structure~\cite{Lee2024,Osumi2024,Fedchenko2024}.

Understanding the origin of altermagnetism in these materials is necessary to study the detailed nature of this phase, and predict new physical properties and useful functionalities for applications. For such purposes, a crucial and useful step is to identify realistic minimal tight-binding models that provide a platform to study the altermagnetic phase and understand what favors this phase over conventional ferromagnetism and antiferromagnetism. Additionally, realistic tight-binding models provide a setup to obtain analytic expressions for the Berry curvature and study the anomalous Hall response.

Minimal models should naturally account for, and give insight into: stable altermagnetic order, the characteristic momentum dependent spin-splittings, and SOC-generated Berry curvature. In addition, these models should be sufficiently general to allow for an understanding of altermagnetism in monoclinic, orthorhombic, tetragonal, hexagonal, and  cubic materials and give rise to $d$-wave, $g$-wave, or $i$-wave altermagnetism. In this work, we provide such minimal models of altermagnetism.  

Our strategy for developing minimal models relies on the relationship between the site symmetry ($S$) of magnetic atoms and the point group symmetry ($P$) of the space group. This is motivated by the realization that in altermagnetic materials $S$ is generically a smaller group than $P$.  This follows because the magnetic sublattice atoms must be related by elements of $P$ \cite{Smejkal2022Sep}, and these elements therefore cannot belong to $S$. This local point group symmetry breaking allows the development of local multipolar moments that are symmetry forbidden in $P$~\cite{Hayami2018}. For example in  RuO$_2$,  the Ru has site symmetry $S=D_{2h}$ while the point group symmetry is $P=D_{4h}$. $S$ allows for local $xy$  quadrupolar order to appear at the two Ru sites. $P$ implies that this $xy$ quadrupolar order is of opposite sign on the two Ru sites. In our minimal models, this local point group symmetry breaking is key to determining the structure of the altermagnetic spin splitting. Since these altermagnetic states are typically inversion invariant, it is natural to consider groups $P$ and $S$ that contain inversion symmetry.

Specifically, we construct models for all space groups that contain inversion symmetry and also contain inversion symmetric Wyckoff positions of multiplicity 2. Our minimal models therefore exhibit two  bands in the paramagnetic state. It is worthwhile contrasting our models with  simpler single-band models~\cite{Oganesyan2001,Wu2007} in which Fermi-surface instabilities of the Pomeranchuk type occur in the spin-triplet channel with high orbital partial waves. In these models, the order parameter is the altermagnetic spin-splitting itself. In our case the altermagnetic spin splitting is a secondary order parameter which is induced through a combination of N\'{e}el order (the primary order parameter) and  the local point group symmetry breaking discussed above. We believe our minimal models are more realistic than  single band models for two reasons: i) there are no known microscopic theories that give rise to altermagnetic Pomeranchuk instabilities; ii) DFT results for altermagnetism in RuO$_2$ show that the largest band splittings in the altermagnetic state occur at band degeneracies \cite{Berlijn2017Feb,Ahn2019}, such degeneracies are not present in single-band Pomeranchuk models.

We note that several earlier works have applied various tight-binding models to address e.g. spin-wave dispersions and superconductivity of altermagnets~\cite{Maier2023,Brekke}. These toy models focused on 2D square and Lieb lattices. Here, we take a different strategy by systematically developing symmetry-dictated minimal models for all centrosymmetric space groups with magnetic atoms occupying inversion symmetric Wyckoff positions with multiplicity two. Through such a comprehensive investigation, we can gain insight into the universal properties of altermagnets, such as the band structure properties favorable for altermagnetism and a general form for the Berry curvature in altermagnets. Additionally, through comparison with DFT, our models allow for material-specific studies, as demonstrated below. Finally, distinct from earlier studies~\cite{Smejkal2020Jun,Fang2023,Fernandes2024}, all minimal models provided below contain the symmetry-allowed momentum-dependent SOC throughout the Brillouin zone (BZ), crucial for capturing important band degeneracies at the faces of the BZ.

The tight-binding models we develop are compared to DFT results for RuO$_2$ to demonstrate that our models capture key properties of the band structure and altermagnetic spin splittings.  By introducing an altermagnetic susceptibility and using susceptibility analyses and selfconsistent Hartree-Fock approaches, we show that these minimal models indeed give rise to altermagnetism for a broad range of parameters. This susceptibility analysis sheds insight into the reason that of the 53 inversion symmetric materials identified as altermagnetic in Ref.~\onlinecite{Guo2023Mar}, 52 belong to  non-symmorphic space groups. In particular, we show that non-symmorphic symmetry-required band degeneracies help stabilize altermagnetism. In addition, we derive the form of the SOC for these models. This SOC is important for understanding the anomalous Hall effect in altermagnets. We provide a general analytic expression for the  SOC-derived
Berry curvature. The resulting Berry curvature is {\it linear} (as opposed to quadratic as found in previous minimal models \cite{Smejkal2020Jun,Fang2023}) in the SOC. This provides a natural explanation for the large crystal Hall effect in altermagnets. Finally, we apply our model to tetragonal, e.g.  MnF$_2$ and RuO$_2$, and orthorhombic, e.g. $\kappa$-Cl and FeSb$_2$, $d$-wave altermagnets, to hexagonal, e.g. CrSb and MnTe, $g$-wave altermagnets, and to cubic $i$-wave altermagnets.

The paper is organized at follows: in Sec.~\ref{sec:minimal_TBH} we expose the minimal model reproducing the DFT band structure for RuO$_2$, which justifies the choice of the altermagnetic order parameter. In Sec.~\ref{sec:susceptibilities}, we focus on the single-orbital minimal model to analyze and compare the bare susceptibilities in the ferromagnetic and the altermagnetic channels. We show that this model is sufficient to capture a leading altermagnetic instability by using both the random phase approximation (RPA) and self-consistent Hartree-Fock calculations. In Sec.~\ref{sec:HallEffect}, we consider the same minimal model and use SOC in order to derive an analytic expression linear in the spin-orbit strength for the Berry curvature in the four-band model. In Sec.~\ref{sec:TBM_AMcandidates} we extend the discussion of minimal models to other altermagnetic candidates with orthorhombic, hexagonal, and cubic lattices. Finally, Section~\ref{sec:discussion} presents our discussion and conclusions.

\section{Minimal models for altermagnetism} \label{sec:minimal_TBH}
\subsection{General Considerations}

Our models apply to all centrosymmetric space groups that contain inversion symmetric Wyckoff positions of multiplicity 2. Tables \ref{tab:spacegroups_mo} and \ref{tab:spacegroups_trhc} give these space groups and the corresponding Wyckoff positions. A key input for our models is the relationship between the point group $P$ and the Wyckoff site symmetry group $S$. Specifically, the primary order parameter we consider is N\'{e}el order on the two Wyckoff positions (perhaps unsurprisingly, we show that this order parameter naturally accounts for the DFT bands in the altermagnetic state).  As discussed in Ref.~\onlinecite{McClarty2023Aug}, this N\'{e}el order transforms as  $\Gamma_N\otimes \Gamma^S_A$, where $\Gamma_N$ is an irreducible representation (IR)  of the point group $G$ and $\Gamma_A^S$ is the axial IR of the spin-rotation group.

The IR $\Gamma_N$ plays a central role in our theory and can be identified from the knowledge of $P$ and $S$ by writing $P=S+hS$ where $h$ is a point group symmetry that switches the two Wyckoff positions. $\Gamma_N$ is then identified as the IR of $P$ that has character $1$ for all elements in $S$ and $-1$ for all elements in $hS$. In Tables \ref{tab:spacegroups_mo} and \ref{tab:spacegroups_trhc}, $\Gamma_N$ is given for each space group and  Wyckoff position. Further, as we show explicitly later, the altermagnetic spin-splitting in all our minimal models is given by $f_{\Gamma_N}(\kv)\vec{\sigma}$  where $\vec{\sigma}$ denotes the N\'{e}el spin direction and $f_{\Gamma_N}(\kv)$ is a momentum dependent function with the same symmetry as $\Gamma_N$. In Tables \ref{tab:spacegroups_mo} and \ref{tab:spacegroups_trhc}, under the column Spin-Splitting, we give representative forms of $f_{\Gamma_N}(\kv)$. We note that while we show the microscopic origin $f_{\Gamma_N}(\kv)$ for electronic models that include non-degenerate orbital IRs of the group $S$, the form of $f_{\Gamma_N}(\kv)$ is also correct for degenerate orbital IRs of the group $S$ provided the site symmetry group $S$ is not broken by a local electronic orbital ordering.

\begin{table}[h]
\caption{Space groups and Wyckoff positions of multiplicty 2 that allow altermagnetism: monoclinic and orthorhombic groups. The notation follows the Bilbao crystallographic server \cite{Aroyo:2006,Aroyo2:2006}. We note that some Wyckoff positions appear with an apparent multiplicity greater than 2, this occurs because  a unit cell larger than the primitive unit cell is conventionally used in these cases. }
\label{tab:spacegroups_mo}
\begin{tabular}{c|c|c|c}
\toprule
SG ($P$) & Wyckoff ($S$) & $\Gamma_N$ &Spin Splitting ($f_{\Gamma_N}(\kv)$)\\ \hline
11 ($C_{2h}$) & 2a-2d ($C_i$) & $B_g$ & $\alpha k_yk_x+\beta k_yk_z$ \\ 
12 ($C_{2h}$)& 4e, 4f ($C_i$)&$B_g$& $\alpha k_yk_x+\beta k_yk_z$ \\
13 ($C_{2h}$) & 2a-2d ($C_i$) & $B_g$ & $\alpha k_yk_x+\beta k_yk_z$  \\
14 ($C_{2h}$) & 2a-2d ($C_i$) & $B_g$ & $\alpha k_yk_x+\beta k_yk_z$ \\
15 ($C_{2h}$) & 2a-2d ($C_i$) & $B_g$ & $\alpha k_yk_x+\beta k_yk_z$  \\
\midrule
49 ($D_{2h}$) & 2a-2d ($C_{2h}$) & $B_{1g}$ & $k_xk_y$  \\
51 ($D_{2h}$) & 2a-2d ($C_{2h}$) & $B_{2g}$ & $k_xk_z$ \\
53 ($D_{2h}$) & 2a-2d ($C_{2h}$) & $B_{3g}$ & $k_yk_z$  \\
55 ($D_{2h}$) & 2a-2d ($C_{2h}$) & $B_{1g}$ & $k_xk_y$  \\
58 ($D_{2h}$) & 2a-2d ($C_{2h}$) & $B_{1g}$ & $k_xk_y$  \\
63 ($D_{2h}$) & 4a,4b ($C_{2h}$) & $B_{3g}$ & $k_yk_z$  \\
64 ($D_{2h}$) & 4a,4b ($C_{2h}$) & $B_{3g}$ & $k_yk_z$  \\
65 ($D_{2h}$) & 4e,4f ($C_{2h}$) & $B_{1g}$ & $k_xk_z$ \\
66 ($D_{2h}$) & 4c-4f ($C_{2h}$) & $B_{1g}$ & $k_xk_y$  \\
67 ($D_{2h}$) & 4c-4f ($C_{2h}$) & $B_{3g}$ & $k_xk_y$  \\
72 ($D_{2h}$) & 4c,4d ($C_{2h}$) & $B_{1g}$ & $k_xk_y$  \\
74 ($D_{2h}$) & 4a,4b ($C_{2h}$) & $B_{3g}$ & $k_xk_y$  \\
74 ($D_{2h}$) & 4c,4d ($C_{2h}$) & $B_{2g}$ & $k_xk_y$ \\

\bottomrule
\end{tabular}
\end{table}

\begin{table}[h]
\caption{Space groups and Wyckoff positions that allow altermagnetism: tetragonal, rhombehedral, hexagonal, and cubic groups.}
\label{tab:spacegroups_trhc}
\begin{tabular}{c|c|c|c}
\toprule
SG ($P$) & Wyckoff ($S$) & $\Gamma_N$ & Spin Splitting ($f_{\Gamma_N}(\kv)$)\\ \hline
83 ($C_{4h}$) & 2e,2f ($C_{2h}$) & $B_{g}$ & $\alpha k_xk_y+\beta(k_x^2-k_y^2)$  \\
84 ($C_{4h}$) & 2a-2d ($C_{2h}$) & $B_{g}$ & $\alpha k_xk_y+\beta(k_x^2-k_y^2)$ \\
87 ($C_{4h}$) & 4c ($C_{2h}$) & $B_{g}$ & $\alpha k_xk_y+\beta(k_x^2-k_y^2)$ \\
\midrule
123 ($D_{4h}$) & 2e,2f ($D_{2h}$) & $B_{1g}$ & $k_x^2-k_y^2$ \\
124 ($D_{4h}$) & 2b,2d ($C_{4h}$) & $A_{2g}$ & $k_xk_y(k_x^2-k_y^2)$  \\
127 ($D_{4h}$) & 2a,2b ($C_{4h}$) & $A_{2g}$ & $k_xk_y(k_x^2-k_y^2)$  \\
127 ($D_{4h}$) & 2c,2d ($D_{2h}$) & $B_{2g}$ & $k_xk_y$ \\
128 ($D_{4h}$) & 2a,2b ($C_{4h}$) & $A_{2g}$ & $k_xk_y(k_x^2-k_y^2)$   \\
131 ($D_{4h}$) & 2a-2d ($D_{2h}$) & $B_{1g}$ & $k_x^2-k_y^2$ \\
132 ($D_{4h}$) & 2a,2c ($D_{2h}$) & $B_{2g}$ & $k_xk_y$  \\
136 ($D_{4h}$) & 2a,2b ($D_{2h}$) & $B_{2g}$ & $k_xk_y$  \\
139 ($D_{4h}$) & 4c ($D_{2h}$) & $B_{1g}$ & $k_x^2-k_y^2$  \\
140 ($D_{4h}$) & 4c ($C_{4h}$) & $A_{2g}$ & $k_xk_y(k_x^2-k_y^2)$   \\
140 ($D_{4h}$) & 4d ($D_{2h}$) & $B_{2g}$ & $k_xk_y$  \\
\midrule
163 ($D_{3d}$) & 2b ($S_6$) & $A_{2g}$ & $k_yk_z(k_y^2-3k_x^2)$   \\
165 ($D_{3d}$) & 2b ($S_6$) & $A_{2g}$ & $k_xk_z(k_x^2-3k_y^2)$  \\
167 ($D_{3d}$) & 6b ($S_6$) & $A_{2g}$ & $k_xk_z(k_x^2-3k_y^2)$  \\
\midrule
176 ($C_{6h}$) & 2b ($S_6$) & $B_{g}$ & $\alpha k_yk_z(k_y^2-3k_x^2)$\\&&&$+\beta k_xk_z(k_x^2-3k_y^2)$  \\
\midrule
192 ($D_{6h}$) & 2b ($C_{6h}$) & $A_{2g}$ & $k_xk_y(k_x^2-3k_y^2)(k_y^2-3k_x^2)$   \\
193 ($D_{6h}$) & 2b ($D_{3d}$) & $B_{2g}$ & $k_xk_z(k_x^2-3k_y^2)$   \\
194 ($D_{6h}$) & 2a ($D_{3d}$) & $B_{1g}$ & $k_yk_z(3k_x^2-k_y^2)$ \\
\midrule
223 ($O_{h}$) & 2a ($D_{3d}$) & $A_{2g}$ & $k_x^4(k_y^2-k_z^2)$\\&&&$+k_y^4(k_z^2-k_x^2)$ \\&&&$+k_z^4(k_x^2-k_y^2)$  \\
\bottomrule
\end{tabular}
\end{table}

\subsection{General Minimal Electronic Model}

To construct our minimal tight-binding models we require $\Gamma_N$ identified in Tables \ref{tab:spacegroups_mo} and \ref{tab:spacegroups_trhc}. For ease of presentation, we restrict ourselves to explicitly providing models for the twenty-seven entries that have primitive unit cells (these are the Wyckoff positions with multiplicity explicitly labeled by 2 in Table \ref{tab:spacegroups_mo} and \ref{tab:spacegroups_trhc}). The corresponding two sublattice positions of the magnetic ions in the unit cell will be labelled by the Pauli matrices $\tau_i$, we will label spin degrees of freedom by Pauli matrices $\sigma_i$. We also assume that the orbital degrees of freedom belong to a singly-degenerate IR of the Wyckoff site symmetry group $S$. It is possible to consider orbitals that belong to degenerate IRs, but these models will necessarily contain more degrees of freedom and hence are not minimal (we carry out a limited investigation of the multi-orbital case for  RuO$_2$ and find that the key features of our minimal model persist). Our minimal model is remarkably versatile: specifically the  model can be  applied to monoclinic, orthorhombic, tetragonal, rhombehedral, hexagonal, and cubic space groups and allows for $d$-wave, $g$-wave, and $i$-wave altermagnetism. Furthermore, this models correctly capture the spin-splittings and the largest band splittings seen by  DFT in the altermagnetic state.

The general minimal model for altermagnetism 
 has the form 
\begin{equation}
    H= \varepsilon_{0,\kv} + t_{x,\kv}\tau_x+t_{z,\kv}\tau_z+\tau_y \Vec{\lambda}_{\kv} \cdot \vec{\sigma} + \tau_z \Vec{J}\cdot \Vec{\sigma},
\label{eq:minimal_general_model}
\end{equation}
with a sublattice independent dispersion  $\varepsilon_{0,\kv}$, inter- and intra-sublattice hopping coefficients $t_{x,\kv}$ and $t_{z,\kv}$, a SOC term $\Vec{\lambda}_{\kv}$, and a primary order parameter $\Vec{J}$. Here the time-reversal symmetry operator is $T=i\tau_0\sigma_y K$ (where $K$ is complex conjugation). The parameters in this minimal model are constrained by the space group, point group $G$, and Wyckoff site symmetry group $S$. We have restricted our minimal models for non-degenerate IRs of $S$, and Eq.~\eqref{eq:minimal_general_model} is valid for all such non-degenerate IRs since the Hamiltonian is built from electronic bilinears that are independent of any sign change that arise for a local rotation.  The sublattice operators $\tau_0$ and $\tau_x$ are invariant under $P$  and the operators $\tau_y$ and $\tau_z$ belong to the IR  $\Gamma_N$ shown in Tables \ref{tab:spacegroups_mo} and \ref{tab:spacegroups_trhc} (this follows because these two operators change sign under the interchange of the two sublattice sites).
In addition, translation symmetry implies $\varepsilon_{0,\kv}=\varepsilon_{0,\kv+\Gv}$ and $t_{z,\kv}=t_{z,\kv+\Gv}$ where $\Gv$ is a reciprocal lattice
vector and, since $\tau_y$ and $\tau_x$ couple the two magnetic atoms, $t_{x,\kv+\Gv}=e^{i\Gv\cdot\tv}t_{x,\kv}$ and   $\Vec{\lambda}_{\kv+\Gv}=e^{i\Gv\cdot \tv}\Vec{\lambda}_{\kv}$ 
 where $\tv$ is the translation between the two magnetic atoms in the unit cell. All coefficients are even under $\kv\rightarrow -\kv$ due to the presence of inversion $I$ in the Wyckoff site symmetry group $S$. 

Prior to providing specific examples and justifying our minimal model with materials examples, we highlight important general properties of the parameters that appear. First we note our altermagnetic order parameter encodes collinear moments parallel to  $\Vec{J}$ that have opposite orientation on the two sublattice magnetic atoms. This term carries no $\kv$ dependence. The intra-sublattice hopping term $t_{z,\kv}$ plays an important role in our theory and its $\kv$ dependence must share the same symmetry as the $\tau_z$ operator - hence $t_{z,\kv}$ belongs to the non-trivial IR $\Gamma_N$.  This term quantifies the existence  of local multipole moments that have opposite sign on the two magnetic atoms and appears due to the local point-group symmetry breaking at the magnetic atom position. As discussed in more detail below, it is this term that gives rise to the momentum-dependent spin splitting that defines altermagnets. The hopping parameter $t_{x,\kv}$ has the full point group symmetry. As we show later, if this hopping parameter is zero, then altermagnetism and ferromagnetism are degenerate within an RPA treatment. Finally, the SOC term follows from $T$ and $I$ symmetries.  In particular, both the spin operators and  $\tau_y$ are odd under $T$, so their product is $T$-invariant. From $I$ symmetry, $\Vec{\lambda}_{\kv}=\Vec{\lambda}_{-\kv}$ and hence this is  also $T$ invariant. As we show below, $\Vec{\lambda}_{\kv}$ gives rise to a Berry curvature that is {\it linear} in the magnitude of SOC.

It is informative to consider the dispersion relation when SOC vanishes. This is given by
\begin{equation}
    E_{\alpha=\pm} =\varepsilon_{0,\kv} + \alpha \Big( t_{x,\kv}^2 + (t_{z,\kv}+ \Vec{J}\cdot \Vec{\sigma})^2\Big)^{1/2}.
    \label{eq:dispersion_minimal_model_noSOC}
\end{equation}
This reveals that the altermagnetic spin-splitting appears through the product $t_{z,\kv}\Vec{J}\cdot \Vec{\sigma}$  
and hence is a consequence of the interplay between the local symmetry breaking and the N\'{e}el order. 
This provides a microscopic realization of the  Ginzburg-Landau bilinear coupling between the N\'{e}el order and  even-parity, odd-time reversal, octupolar order, which gives rise to the altermagnetic spin-splitting \cite{Hayami2018}, as discussed in the context of RuO$_2$ in
Ref.~\onlinecite{McClarty2023Aug}.

\begin{figure}[t]
\begin{center}
\includegraphics[angle=0,width=.95\linewidth]{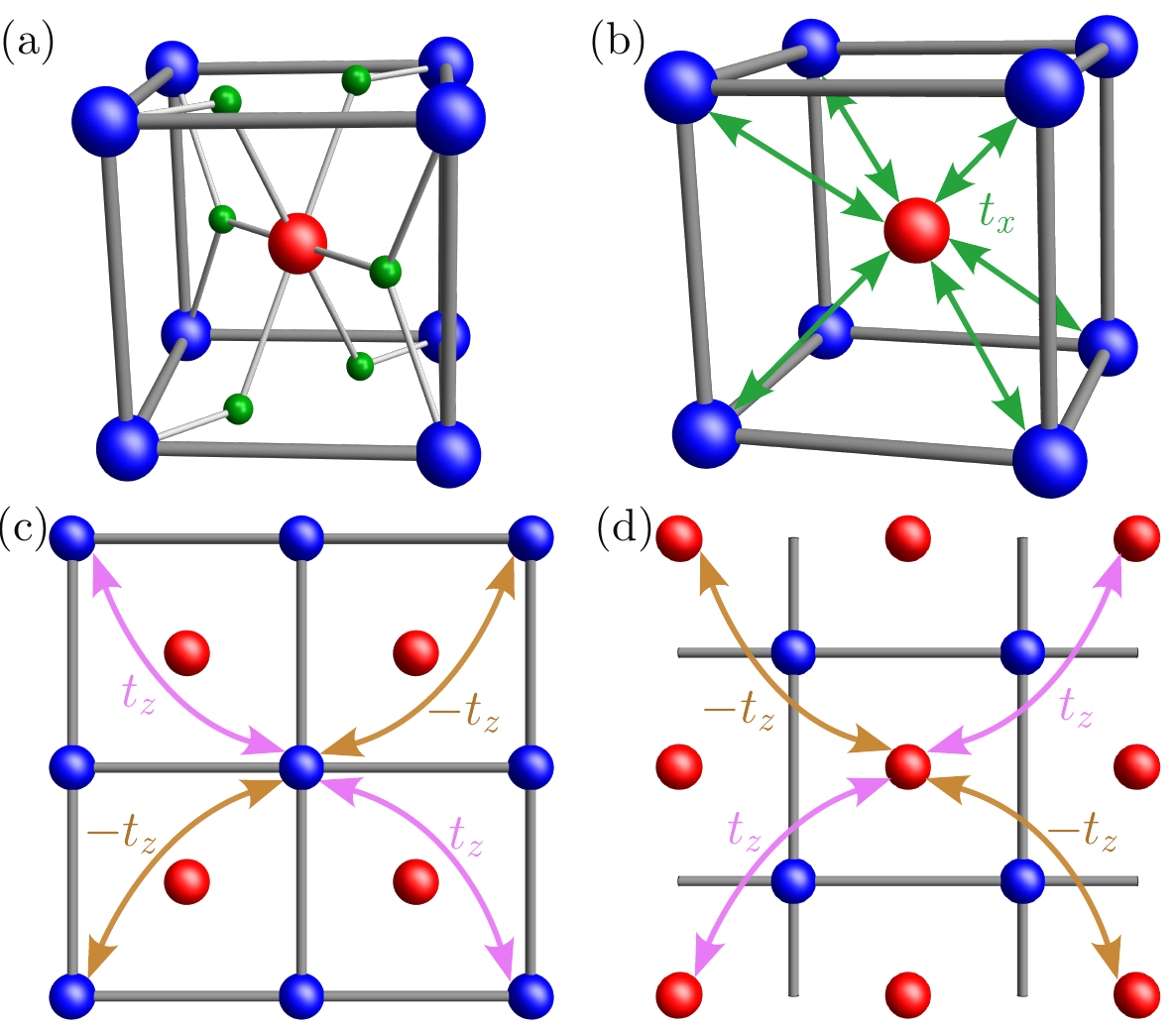}
\caption{Sketch of the crystal structure for the tetragonal SG 136 and the relevant hoppings in the minimal model presented in Eq.~\eqref{eq:minimal_general_model}, with the red and blue colors representing the two sublattices. (a) Crystal structure including the non-magnetic atoms denoted by the green color. (b) Three-dimensional lattice illustration of the $t_{x,\kv}$ hopping between sublattices. (c),(d) Top view of the lattice showing the $t_{z,\kv}$ hopping with the symmetry-imposed sign change, which has an opposite sign on the two sublattices due to the presence of the non-magnetic atoms.}
\label{fig:sketch_hoppings}
\end{center}
\end{figure}

As a specific example of our minimal model, it is worthwhile considering the simplest model for SG 136 and Wyckoff position 2a (this is the Wyckoff position for the magnetic atoms in RuO$_2$, MnF$_2$, NiF$_2$, and CoF$_2$). SG 136 has point group $D_{4h}$ and the 2a Wyckoff position has site symmetry $D_{2h}$, this implies that $\tau_y$ and $ \tau_z$ belong to the $B_{2g}$ representation of $D_{4h}$.
The hoppings $t_{x,\kv}$ and $t_{z,\kv}$ are illustrated in Fig.~\ref{fig:sketch_hoppings}. Here,  $t_{x,\kv}=t_0\cos \frac{k_x}{2} \cos \frac{k_y}{2} \cos\frac{k_z}{2}$, the factors $\frac{k_i}{2}$ appear because of the condition  $t_{x,\kv+\Gv}=e^{i\Gv\cdot\tv}t_{x,\kv}$ with $\tv=(\frac{1}{2},\frac{1}{2},\frac{1}{2})$. In addition, $t_{z,\kv}=t_{z0}\sin k_x \sin k_y$ has $d$-wave symmetry-imposed sign changes which follow from the condition that $\tau_z$ belongs to the $B_{2g}$ representation and quantifies local point group symmetry breaking through the appearance of $xy$ quadrupolar order on the Ru sites due to surrounding O atoms.
We note that for $k_x=\pi$ or $k_y=\pi$ both $t_{x,\kv}$ and $t_{z,\kv}$ vanish, revealing symmetry-required nodal planes that exist due to the non-symmorphic symmetry elements $\{C_{2x}|\frac{1}{2},\frac{1}{2},\frac{1}{2}\}$
and $\{C_{2y}|\frac{1}{2},\frac{1}{2},\frac{1}{2}\}$ \cite{Suh:2023}. As we show later, these nodal planes aid in stabilizing the altermagnetic state.
Further, for the spin degrees of freedom, $\sigma_z \sim A_{2g}$ and $(\sigma_x, \sigma_y) \sim E_g$.
Consequently, symmetry arguments imply that $\Vec{\lambda}_{\kv}$ is given by 
\begin{equation}
    \begin{aligned}
        & \lambda_{x,\kv} = \lambda \sin \frac{k_z}{2} \sin \frac{k_x}{2} \cos \frac{k_y}{2}, \\
        & \lambda_{y,\kv} = - \lambda \sin \frac{k_z}{2} \sin \frac{k_y}{2} \cos \frac{k_x}{2}, \\
        & \lambda_{z,\kv} = \lambda_z \cos \frac{k_z}{2} \cos \frac{k_x}{2} \cos \frac{k_y}{2} (\cos k_x - \cos k_y).
    \end{aligned}
    \label{eq:xyorb_SOC}
\end{equation}
This approach has been applied to define the simplest tight-binding form for the parameters $t_{x,\kv},t_{z,\kv},\vec{\lambda}_{\kv}$ for 27 Wyckoff positions in Appendix~\ref{ap:otherspacegroups}, providing many altermagnetic examples.
We note that in the following we will consider additional hopping terms in our minimal model to fit DFT results. However, for SG 136, $\Vec{\lambda}_{\kv}$ remains the same. 

\subsection{Dispersion relations, altermagnetic spin-splittings, and Weyl-lines}

The general form of the dispersion for the minimal model in Eq.~\eqref{eq:minimal_general_model} is given by
\begin{equation}
    \begin{aligned}
    E_{\alpha=\pm,\beta=\pm} = & \, \varepsilon_{0,\kv} + \alpha \Big( t_{x,\kv}^2 + t_{z,\kv}^2 + \Vec{\lambda}_{\kv}^2 + \Vec{J}^2 \\ 
    &+ \beta 2\sqrt{t_{z,\kv}^2 \Vec{J}^2 + (\Vec{\lambda}_\kv \cross \Vec{J})^2}\Big)^{1/2}.
    \end{aligned}
    \label{eq:dispersion_minimal_model}
\end{equation}
In the limit of vanishing SOC, $\Vec{\lambda}_\kv=0$, the four bands are generally non-degenerate, except in two cases. The first case occurs when $t_{z,\kv}=0$. This defines Weyl planes in momentum space with two two-fold degenerate bands where the spin-splitting vanishes. These Weyl planes are symmetry imposed and always present due to the non-trivial symmetry of operator $\tau_z$. These Weyl planes are the nodes of the usual altermagnetic spin-splitting and, as mentioned earlier, are entirely given here by the vanishing of the intra-sublatice hopping $t_{z,\kv}$.
The second case corresponds to $t_{x,\kv}=0$ (which is often required by symmetry to occur on the BZ boundary) and $t_{z,\kv} = \pm |\Vec{J}|$, which defines Weyl lines. 
Along these lines, we have a two-fold degeneracy together with two non-degenerate bands. These lines are not symmetry imposed and  appear when $|t_{z,\kv}|$ is larger than $|\Vec{J}|$. As we will discuss in Sec.~\ref{sec:HallEffect}, this occurs in RuO$_2$ and these Weyl lines can be important for the Berry curvature.

When $\Vec{\lambda}_{\kv}\neq 0$, the Weyl planes discussed above (for which $t_{z,\kv}=0$) become partially gapped to form Weyl lines or Weyl points.  Specifically, when $t_{z,\kv}=0$, Weyl lines occur when symmetry requires the cross product between $\Vec{\lambda}_\kv$ and $\Vec{J}$ to vanish, $\Vec{\lambda}_\kv \cross \Vec{J}=0$. The Weyl lines discussed in the previous paragraph become gapped if $\Vec{\lambda}_\kv \cdot \Vec{J} \ne 0$ and survive otherwise. Related Weyl lines have been discussed in Refs.~\cite{Fernandes2024,yu2023,Antonenko2024Feb,Attias2024Feb}.

\subsection{Application to \texorpdfstring{RuO$_2$}{RuO2}: justification for the order parameter}

Here we demonstrate how this model can describe the non-magnetic band structure of tetragonal altermagnetic material candidate RuO$_2$ found in DFT~\cite{Ahn2019}, and captures the altermagnetic spin-splitting of the bands. We are giving explicit models for materials and examine these with realistic parameters, thus all energies are in units of eV (unless specified otherwise).
As shown in Appendix \ref{ap:orbital_projection}, the orbital projection of the non-relativistic DFT bands reveals that the $d_{xy},d_{xz},d_{yz}$ orbitals form bands crossing the Fermi level. We show that the minimal model can be generalized to a multi-orbital case since the order parameter describing the spin splitting comes only from opposite spins in the two sublattices. In Appendix~\ref{ap:DFT_MnF2} we discuss the case of another tetragonal material candidate, MnF$_2$, where DFT has also identified an altermagnetic phase~\cite{Yuan2020Jul}.

To construct the specific one-orbital tight-binding models for this tetragonal material, we consider $d_{xy}$ orbitals on the Wyckoff position 2a for which the example discussed in Sec.~\ref{sec:minimal_TBH}A applies. Specifically, we take the dispersion
\begin{equation}
    \begin{aligned}
        \varepsilon_{0,\kv} = & \; t_1(\cos k_x + \cos k_y) - \mu + t_2 \cos k_z + t_3 \cos k_x \cos k_y \\
        & + t_4 (\cos k_x + \cos k_y) \cos k_z  + t_5 \cos k_x \cos k_y \cos k_z,
    \end{aligned}
    \label{eq:xyorb_dispersion}
\end{equation}
the hoppings
\begin{equation}
    \begin{aligned}
        & t_{x,\kv} =  t_8 \cos \frac{k_x}{2} \cos \frac{k_y}{2} \cos\frac{k_z}{2},\\
        & t_{z,\kv} = t_6 \sin k_x \sin k_y  + t_7 \sin k_x \sin k_y \cos k_z,
    \end{aligned}
    \label{eq:xyorb_hoppings}
\end{equation}
and the SOC terms given by Eq.~(\ref{eq:xyorb_SOC}).

\begin{figure}[!t]
\begin{center}
\includegraphics[angle=0,width=\linewidth]{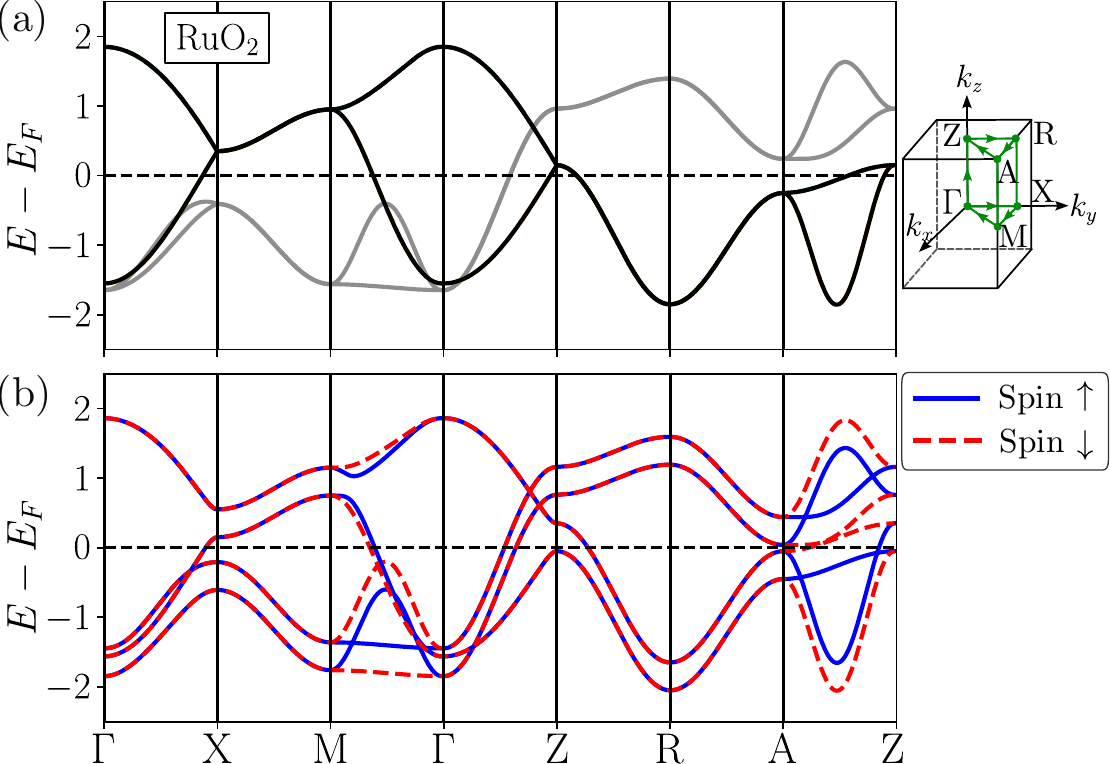}
\caption{Normal state (a) and altermagnetic (b) band structures for RuO$_2$ obtained from the minimal model in Eq.~\eqref{eq:minimal_general_model} taking Eqs.~\eqref{eq:xyorb_dispersion}-\eqref{eq:xyorb_hoppings}, with hopping parameters detailed in Appendix~\ref{ap:hoppings} to reproduce the DFT results (see Appendix~\ref{ap:orbital_projection} and Ref.~\cite{Ahn2019}) and $J_z = 0.2$ in (b). The gray bands correspond to the two-orbital model in Eqs.~\eqref{eq:3DTBM_xzyzorb}-\eqref{eq:hoppings_2orb} with the hoppings in Appendix~\ref{ap:hoppings}. From the latter model there are two more bands with a two-fold spin-degeneracy higher in energy (not shown).
}
\label{fig:RuO2_bands}
\end{center}
\end{figure}

Figure~\ref{fig:RuO2_bands}(a) displays the normal state bands obtained from this minimal model using an appropriate choice of hopping parameters specified in Appendix \ref{ap:hoppings}. Notably, the minimal model reproduces the main features of the bands, capturing the crossings at the Fermi level, as well as the characteristic nodal lines along the X-M and Z-R-A directions. 
The $t_{z,\kv}\tau_z$ term in the minimal model in Eq.~\eqref{eq:minimal_general_model} is crucial since it is the only one controlling the splitting of the bands in the A-Z line. In addition, the $t_{x,\kv}\tau_x$ term is responsible for the splitting of the band degeneracies in the $\Gamma$-X and M-$\Gamma$-Z directions. 
The $\kv$-dependent SOC terms in Eq.~\eqref{eq:xyorb_SOC} open a band splitting along the Z-R-A lines (not shown), in agreement with the relativistic DFT results included in Appendix~\ref{ap:orbital_projection}.
Figure~\ref{fig:RuO2_bands}(b) shows the band structure in the altermagnetic state obtained by including an order parameter as described in Eq.~\eqref{eq:minimal_general_model}, resulting in spin splittings in agreement with those obtained by the magnetic DFT results~\cite{Ahn2019}. Specifically, this model reproduces the altermagnetic spin splitting along the M-$\Gamma$ and A-Z directions, driven solely by the term $t_{z,\kv}\tau_z$.

The coupled $d_{xz}/d_{yz}$ orbitals also cross the Fermi level in the RuO$_2$ DFT bands (see Appendix~\ref{ap:orbital_projection}). Therefore, we have additionally produced a two-orbital tight-binding model without including couplings with the one-orbital model.
For the two-orbital model, we duplicate the terms of the minimal model in Eq.~\eqref{eq:minimal_general_model} for the two orbitals and supplement it with the following symmetry-allowed terms,
\begin{equation}
    \begin{aligned}
        &\gamma_z (t_{a,\kv} {+} t_{b,\kv} \tau_x {+} t_{c,\kv} \tau_z) + \gamma_x (t_{d,\kv}{+}t_{e,\kv}\tau_x {+} t_{f,\kv} \tau_z) \\
        &+ t_{g,\kv} \gamma_y \tau_y + \lambda_0 \gamma_y \tau_0 \sigma_z,
    \end{aligned}
    \label{eq:3DTBM_xzyzorb}
\end{equation}
with 
\begin{equation}
    \begin{aligned}
        & t_{a,\kv} = (t_9 + t_{10} \cos k_z )(\cos k_x - \cos k_y), \\
        & t_{b,\kv} = t_{11} \cos \frac{k_x}{2}\cos\frac{k_y}{2} \cos \frac{k_z}{2} (\cos k_x - \cos k_y), \\
        & t_{c,\kv} =  t_{12} \sin k_x \sin k_y (\cos k_x - \cos k_y), \\
        & t_{d,\kv} = t_{13} \sin k_x \sin k_y, \\
        & t_{e,\kv} = t_{14} \sin\frac{k_x}{2} \sin\frac{k_y}{2} \cos\frac{k_z}{2}, \\
        & t_{f,\kv} =  a_0 + t_{15} (\cos k_x + \cos k_y), \\
        & t_{g,\kv} = t_{16} \cos\frac{k_x}{2} \cos\frac{k_y}{2} \cos\frac{k_z}{2} (\cos k_x - \cos k_y), \\
        \label{eq:hoppings_2orb}
    \end{aligned}
\end{equation}
where we have introduced the Pauli matrices $\gamma_i$ to represent orbital space, with $\gamma_x \sim B_{2g}$, $\gamma_y \sim A_{2g}$ and $\gamma_z \sim B_{1g}$ in the point group $D_{4h}$. In contrast to the previous case, the two-orbital model contains a symmetry-allowed on-site SOC term $\lambda_0$, which splits the bands at the $\Gamma$ point. Note that $t_{13} \sin k_x \sin k_y \gamma_x$ has the same symmetry as $t_{z,\kv}$ and therefore also controls the spin splitting along M-$\Gamma$ and A-Z lines. The band structure from the $d_{xz}/d_{yz}$ orbitals is also included in Fig.~\ref{fig:RuO2_bands}, and reproduces the relevant features of the DFT bands demonstrating that our choice of order parameter provides an accurate description of the altermagnetic state.

\section{Susceptibilities and stabilization of altermagnetism} \label{sec:susceptibilities}

In the previous section, we introduced general minimal models describing altermagnetism. The purpose of the current section is to demonstrate that these models indeed give rise to a leading altermagnetic instability and to examine the mechanism driving altermagnetism. For simplicity, we focus on the one-orbital model case shown in Eq.~\eqref{eq:minimal_general_model}, even though the discussion can also be extended to the multi-orbital case. In addition, in order to analyze the expressions for the susceptibility and describe simple mechanisms stabilizing altermagnetism, we neglect SOC in this section. We return to the role of SOC in Sec.~\ref{sec:HallEffect} when discussing Berry curvature and the altermagnetic driven anomalous Hall effect.

\subsection{Analytic expressions of the susceptibilities}\label{sec:analytic_bare_suscept}
To gain insight into what determines a leading altermagnetic versus ferromagnetic instability, we have obtained analytic expressions for the bare susceptibility in band space considering the minimal Hamiltonian
\begin{equation}
    {H'} = \varepsilon_{0,\kv} + t_{x,\kv} \tau_x + t_{z,\kv} \tau_z.
\end{equation}
From this expression we see that the unitary transformation from sublattice to band basis is generally $\kv-$dependent.
The transformation matrix
\begin{equation}
    U_\kv = \begin{pmatrix}
        \cos \frac{\theta_\kv}{2} & \sin \frac{\theta_\kv}{2} \\
        - \sin \frac{\theta_\kv}{2} & \cos \frac{\theta_\kv}{2}
    \end{pmatrix}
    \label{eq:transf_bandbasis}
\end{equation}
diagonalizes $H'$, i.e. $U_\kv^\dagger H' U_\kv=\mathrm{diag}(E^+_{\kv},E^-_{\kv})$, where $\cos \theta_\kv = \frac{t_{z,\kv}}{\sqrt{t_{z,\kv}^2 + t_{x,\kv}^2}}$ and $\sin \theta_\kv = \frac{t_{x,\kv}}{\sqrt{t_{z,\kv}^2 + t_{x,\kv}^2}}$.  

The susceptibility in the usual spin channel is
\begin{equation}
    \chi^{\textrm{FM}} (\qv, iq_n) = - \int_0^\beta e^{iq_n\tau} \langle T_\tau S_\qv (\tau) S_{-\qv} (0) \rangle,
\end{equation}
where $S_\qv =  \frac{1}{N} \sum_\kv \Psi^\dagger_{\kv+\qv} \tau_0 \Psi_\kv$, with the spinor $\Psi_{\kv} = (\psi_{\kv,1} \ \psi_{\kv,2})^T$ in the sublattice basis. We refer to this as the ferromagnetic channel since this susceptibility diverges at $\qv \rightarrow 0$ close to a ferromagnetic instability.
Transforming to the band basis,
\begin{equation}
    \chi^{\textrm{FM}}(\qv,iq_n) = - \frac{1}{N} \sum_{\kv,a,b} \abs{\bra{u^a_\kv}  \ket{u^b_{\kv + \qv}}}^2  \frac{f(E_\kv^a) - f(E_{\kv + \qv}^b)}{iq_n + E_\kv^a - E_{\kv + \qv}^b}.
    \label{eq:chi_FM_analyticbandbasis}
\end{equation}
Focusing on the uniform static susceptibility, the ferromagnetic spin susceptibility has only intraband terms,
\begin{equation}
    \chi^{\textrm{FM}}(0) = - \frac{1}{N} \sum_\kv \left\{ \frac{df(\varepsilon)}{d\varepsilon}\Bigg|_{\varepsilon = E_\kv^+} + \frac{df(\varepsilon)}{d\varepsilon}\Bigg|_{\varepsilon = E_\kv^-} \right\}.
\end{equation}

Motivated by the RuO$_2$ band structure and the form of the altermagnetic order parameter shown in Eq.~\eqref{eq:minimal_general_model}, we can obtain an equivalent expression for the altermagnetic susceptibility,
\begin{equation}
    \chi^{\textrm{AM}} (\qv, iq_n) = - \int_0^\beta e^{iq_n\tau} \langle T_\tau \Tilde{S}_\qv (\tau) \Tilde{S}_{-\qv} (0) \rangle,
\end{equation}
where now $\Tilde{S}_\qv = \frac{1}{N} \sum_\kv \Psi^\dagger_{\kv+\qv} \tau_z \Psi_\kv$.
In the band basis, the altermagnetic susceptibility becomes 
\begin{equation}
    \chi^{\textrm{AM}}(\qv,iq_n) {=} {-} \frac{1}{N} \sum_{\kv,a,b} \abs{\bra{u^a_\kv}  \tau_z \ket{u^b_{\kv + \qv}}}^2  \frac{f(E_\kv^a) {-} f(E_{\kv + \qv}^b)}{iq_n {+} E_\kv^a {-} E_{\kv + \qv}^b}.
\end{equation}
In the $\qv \rightarrow 0, iq_n \rightarrow 0$ limit, projecting the $\tau_z$ operator onto the band basis using Eq.~\eqref{eq:transf_bandbasis},
\begin{align}
        \chi^{\textrm{AM}}(0) &= \chi^{\textrm{FM}}(0) - \frac{1}{N} \sum_\kv \sin^2 \theta_\kv \Bigg\{ \frac{2[f(E_\kv^-) - f(E_{\kv}^+)]}{E_\kv^- - E_\kv^+} \nonumber \\
        & -  \Bigg[ \frac{df(\varepsilon)}{d\varepsilon}\Bigg|_{\varepsilon = E_\kv^+} + \frac{df(\varepsilon)}{d\varepsilon}\Bigg|_{\varepsilon = E_\kv^-} \Bigg]  \Bigg\}. 
   \label{eq:AM_suscept}
\end{align}
In contrast to the ferromagnetic channel, the altermagnetic susceptibility contains both intraband and interband contributions, and the competition between them determines the leading instability. In order to stabilize altermagnetism, the interband contribution should be larger, whereas if the intraband part is dominant ferromagnetism is leading. Note that $\sin^2 \theta_\kv>0$ is also needed, i.e. a finite $t_{x,\kv}$ term in Eq.~\eqref{eq:minimal_general_model}, since otherwise the two instabilities are degenerate. 

Importantly, Eq.~\eqref{eq:AM_suscept} also shows that band degeneracies enhance the interband susceptibility, as they correspond to $E_\kv^+ - E_\kv^- \rightarrow 0$. Hence, these have an important role in stabilizing altermagnetism. Without considering SOC, the non-symmorphic symmetry ensures these band degeneracies on nodal planes. In the case of RuO$_2$, there are nodal planes in the $x,y$ faces of the BZ and two cross lines on the $z$ face, as seen in Fig.~\ref{fig:RuO2_bands}(a). Altermagnetism is also favored if there exists a nesting line between the two bands, leading to a divergent interband susceptibility. In a three-dimensional picture, a nesting line in momentum space can exist between two spherical Fermi surfaces centered around the same point, giving rise to a cusp in the density of states.

Examining the intra- and interband susceptibilities using the tight-binding Hamiltonian in Eq.~\eqref{eq:minimal_general_model} shows that the band splitting due to the $t_{x,\kv} \tau_x$ term has to be sufficiently large for the interband term to dominate in Eq.~\eqref{eq:AM_suscept}, which is typically the case since $t_{x,\kv}$ corresponds to nearest neighbor hopping, as shown in Fig.~\ref{fig:sketch_hoppings}. Expanding Eq.~\eqref{eq:AM_suscept} close to the A-point reveals that the contribution from the van Hove singularity (see Fig.~\ref{fig:RuO2_bands}) gives rise to a dominant interband susceptibility, thus stabilizing altermagnetism.

\begin{figure}[t]
    \begin{center}
		\includegraphics[angle=0,width=0.95\linewidth]{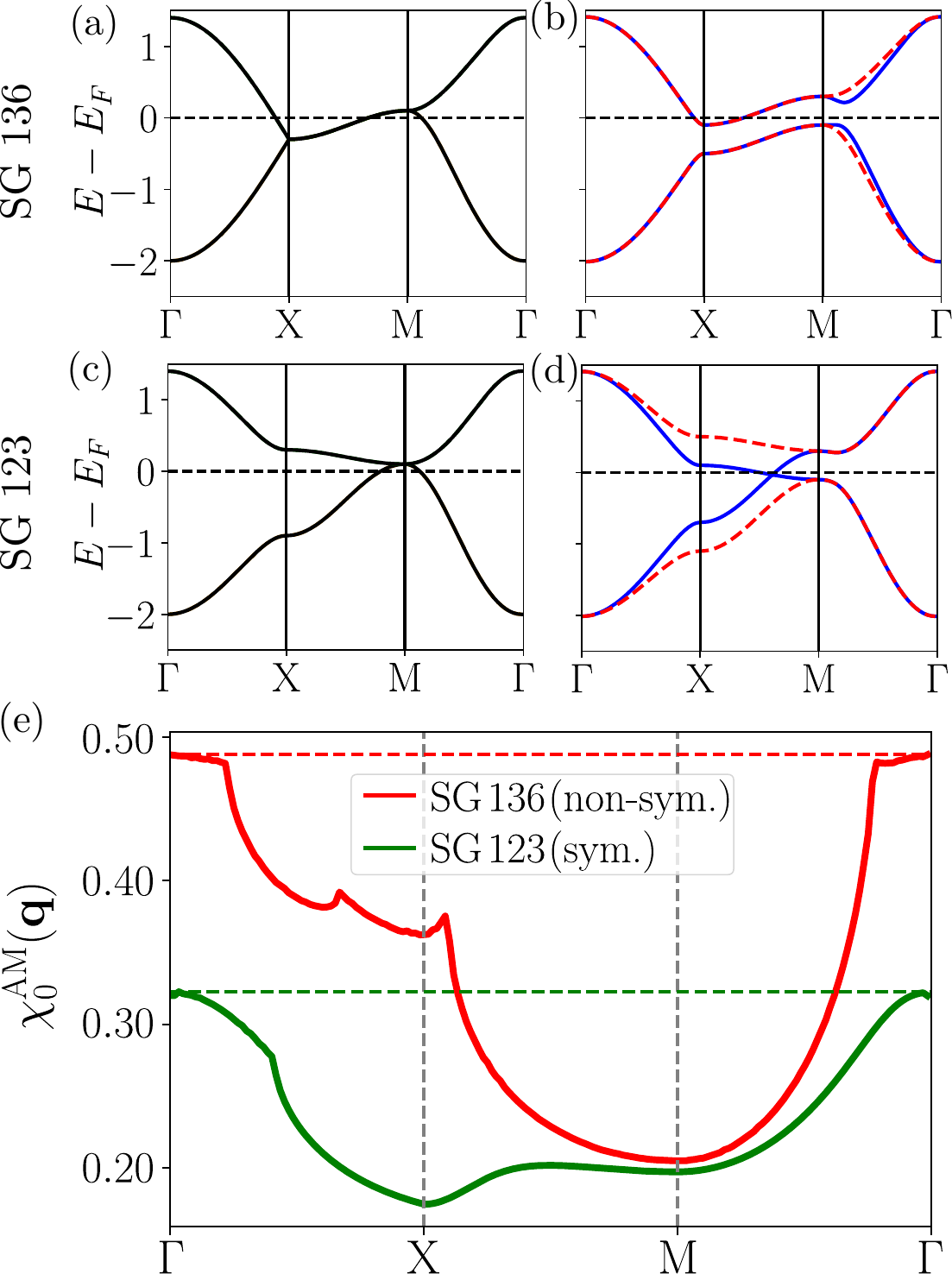}
		\caption{(a,c) Normal state and (b,d) altermagnetic band structures for space groups SG 136 (non-symmorphic) and SG 123 (symmorphic). In both cases, $\{t_1,t_2,t_3,t_4,\mu\} = \{-0.1, 0.1,1.7,0.3,0.2\}$ (see Eqs.~\eqref{eq:2D_minimalmodel_SG136}-\eqref{eq:2D_minimalmodel_SG123}), with $J_z = 0.2$ in (b),(d). (e) Comparison of the altermagnetic susceptibilities, with $T=10^{-4}$ and $n_k=1200^2$, displaying a significantly larger altermagnetic susceptibility in the non-symmorphic case.}
    \label{fig:SG123vs136}	
     \end{center}
\end{figure}
To further emphasize the difference between symmorphic and non-symmorphic space groups, and elucidate the role of band degeneracies in driving the altermagnetic instability, we have compared the band structures and the altermagnetic susceptibilities for a 2D minimal model for SG 123 (symmorphic) and SG 136 (non-symmorphic). SG 136 with Wyckoff position 2a has been discussed in Sec.~\ref{sec:minimal_TBH}, and based on the 2D model discussed in Appendix~\ref{ap:2D_tetragonal_toymodel} inspired by the RuO$_2$ bands in Fig.~\ref{fig:RuO2_bands}(a), we consider the minimal model
\begin{equation}
    \begin{aligned}
    H_{\rm{2D}}^{\rm{SG}136} & = t_1 (\cos k_x + \cos k_y) + t_2 \cos k_x \cos k_y - \mu \\
    &+ t_3 \cos \frac{k_x}{2}\cos \frac{k_y}{2} \tau_x + t_4 \sin k_x \sin k_y \tau_z,
    \end{aligned}
    \label{eq:2D_minimalmodel_SG136}
\end{equation}
with the hoppings illustrated in Fig.~\ref{fig:Ap_2D_minmodel}. 
As a symmorphic example, we consider SG 123 and Wyckoff position 2e, which corresponds to $(0,1/2)$ and $(1/2,0)$ in a 2D model. SG 123 has also point group $D_{4h}$ and the 2e Wyckoff position has site symmetry $D_{2h}$, and thus $\tau_y$ and $\tau_z$ belong to the $B_{1g}$ representation of $D_{4h}$. The minimal model is then given by
\begin{equation}
    \begin{aligned}
    H_{\rm{2D}}^{\rm{SG}123} & = t_1 (\cos k_x + \cos k_y) + t_2 \cos k_x \cos k_y - \mu \\
    &+ t_3 \cos \frac{k_x}{2}\cos \frac{k_y}{2} \tau_x + t_4 (\cos k_x - \cos k_y) \tau_z.
    \end{aligned}
    \label{eq:2D_minimalmodel_SG123}
\end{equation}
Importantly, as opposed to the previous non-symmorphic example, the term $t_{z,\kv} = t_4 (\cos k_x - \cos k_y)$ splits the bands at the BZ boundary.

In Fig.~\ref{fig:SG123vs136}(a) and Fig.~\ref{fig:SG123vs136}(c) we show the normal state band structure for SG 136 and SG 123, respectively. The non-symmorphic space group (see Fig.~\ref{fig:SG123vs136}(a)) features band degeneracies at the BZ boundary, as seen along the X-M direction. On the contrary, the symmorphic space group is crucially different, since the band is only degenerate at the M-point. The altermagnetic band structures are displayed in Fig.~\ref{fig:SG123vs136}(b),(d), showing the vastly distinct altermagnetic spin splitting between the two cases due to the different symmetries of the $t_{z,\kv}$ term. We also note that the required Hubbard interaction to acquire this spin splitting (in mean field theory) is $U=1.017$ for SG 136 while for SG 123 a significantly larger $U=1.72$ would be needed.

Figure~\ref{fig:SG123vs136}(e) shows the non-interacting altermagnetic susceptibility for both space groups. As seen, for SG 136 the altermagnetic susceptibility is notably larger when compared to SG 123. To understand this result, Eq.~\eqref{eq:AM_suscept} is crucial as it points out that band degeneracies have an important role in stabilizing altermagnetism. Hence, in terms of altermagnetism the main difference between symmorphic and non-symmorphic space groups is indeed the symmetry-imposed band degeneracies characteristic of non-symmorphic space groups.

\subsection{Stabilization of altermagnetism}\label{subsec:RPA}

To demonstrate that the minimal tight-binding models indeed give rise to a leading altermagnetic instability, we start from the Hamiltonian $H'$ and consider standard intra-orbital Hubbard interaction $U$ given by
\begin{equation}
    H_{\rm int} = U \sum_{i,\mu} n_{i,\mu, \uparrow} n_{i,\mu, \downarrow},
    \label{eq_hubbard}
\end{equation}
where $\mu$ denotes the sublattice index, and show that this is sufficient to give rise to altermagnetism. Note that Ref.~\onlinecite{Yu2024} demonstrated that Eq.~(\ref{eq_hubbard}) gives rise to the relevant altermagnetic interaction in the band basis in the case of coincident van Hove singularities.

In the multi-orbital RPA approximation, the RPA susceptibility matrix can be written as~\cite{Graser2009,Kreisel2017,Astrid2019,Maier2023}
\begin{equation}
    [\chi_{\rm RPA} (\qv,iq_n)]^{\mu_1,\mu_2}_{\mu_3,\mu_4} = \Big[ \chi_0 (\qv,iq_n)\big(\mathds{1}{-} U \chi_0 (\qv,iq_n)\big)^{{-}1}\Big]^{\mu_1,\mu_2}_{\mu_3,\mu_4},
    \label{eq:multiorb_RPA}
\end{equation}
where $iq_n$ is a bosonic Matsubara frequency, $[U]^{\mu_1,\mu_2}_{\mu_3,\mu_4} = U$ for $\mu_1 = \mu_2 = \mu_3 = \mu_4$ and the bare susceptibility matrix is given by
\begin{align}
         \relax [\chi_0 (\qv,iq_n)]^{\mu_1,\mu_2}_{\mu_3,\mu_4} =  - \frac{1}{N \beta} \sum_{\kv,i\omega_n}& G^0_{\mu_1 \mu_3} (\kv + \qv, i\omega_n + iq_n) \notag\\
         & \times G^0_{\mu_2 \mu_4} (\kv, i\omega_n).
         \label{eq:bare_suscept_sublat}
\end{align}
The matrix product is calculated by combining the sublattice indices to construct the matrix elements, $A_{\mu_1 + \mu_2 N_s,\mu_3 + \mu_4 N_s}$, with $A=\{\chi_0 (\qv,iq_n), U\}$, where in this case $N_s=2$ since we only have two sublattices, giving thus $4\cross 4$ matrices.
In Eq.~\eqref{eq:bare_suscept_sublat}, the bare Green's function corresponds to 
\begin{equation}
    G^0_{\mu \nu} (\kv, i\omega_n) = \sum_m \frac{u^\mu_m (\kv) u_m^{\nu *} (\kv)}{i\omega_n - E_m (\kv)},
\end{equation}
with $u_m^\mu (\kv)$ the $\kv$-dependent eigenvector connecting band space ($m$) with sublattice space ($\mu$), and $E_m (\kv)$ the corresponding energy eigenvalue of the Hamiltonian $H'$ for band $m$. 

To obtain the physical spin susceptibility in the ferromagnetic channel we sum over both sublattices
\begin{equation}
    \chi^{\rm FM}_{\rm RPA}(\qv,\omega) = \sum_{\mu,\nu} [\chi_{\rm RPA}(\qv,iq_n \rightarrow \omega + i \eta)]^{\mu,\mu}_{\nu,\nu}.
    \label{eq:RPA_FM}
\end{equation}
In analogy with the analysis of the bare susceptibilities in the previous section, we can obtain an expression for the susceptibility in the $\tau_z$ channel. Therefore, the susceptibility in the altermagnetic channel for a two-band model can be written as
\begin{equation}
    \chi^{\rm AM}_{\rm RPA}(\qv,\omega) = \sum_{\mu,\nu} ({-}1)^\mu ({-}1)^\nu [\chi_{\rm RPA} (\qv,iq_n \rightarrow \omega {+} i \eta)]^{\mu,\mu}_{\nu,\nu}.
    \label{eq:RPA_AM}
\end{equation}
If this susceptibility diverges before the usual spin channel, it signifies that the altermagnetic phase is favored over ferromagnetism or spin-density wave order. The form of $\chi^{\rm AM}_{\rm RPA}(\qv,\omega)$ also reveals that the altermagnetic instability is favored when the inter-sublattice components become negative, as shown in Appendix~\ref{ap:suscept_sublatcomp}.

\begin{figure}[t]
\begin{center}
\includegraphics[width=0.95\linewidth]{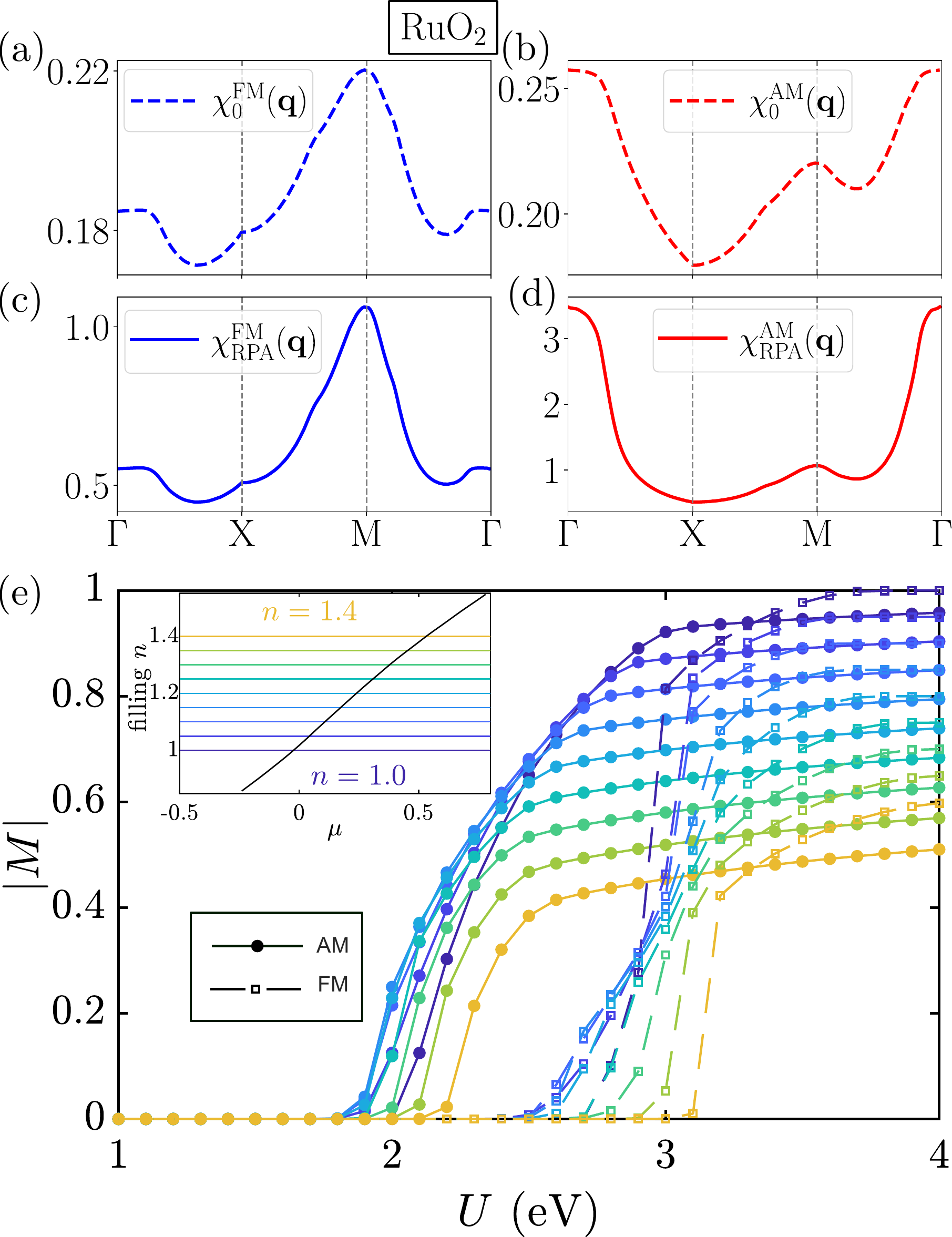}
\caption{(a)-(d) Bare and RPA susceptibilities in the ferromagnetic and altermagnetic channels (see Eqs.~\eqref{eq:RPA_FM},\eqref{eq:RPA_AM}) for RuO$_2$, considering the minimal one-orbital model band structures shown in Fig.~\ref{fig:RuO2_bands}(a), for  $U=1.8$, $T=0.02$ and $n_k=60^3$.
(e) Order parameter $|M|=\sum_{\alpha}| n_{\uparrow,\alpha}-n_{\downarrow,\alpha}|$ for RuO$_2$ from a selfconsistent Hartree-Fock calculation at different fillings $n$ where both an altermagnetic and ferromagnetic order parameter can be stabilized, for $T=0.02$ and $n_k=40^3$.}
\label{fig:HF_RuO2}
\end{center}
\end{figure}

Figure~\ref{fig:HF_RuO2}(a)-(d) display the RPA results for the RuO$_2$ bands shown in Fig.~\ref{fig:RuO2_bands}(a), considering only the one-orbital minimal model in Eq.~\eqref{eq:minimal_general_model}. As seen, altermagnetism becomes the leading instability and diverges at $\qv \rightarrow 0$. We notice that the bare susceptibility exhibits the same momentum structure as the RPA susceptibility, as expected from a single orbital model. Therefore, the discussion in Sec.~\ref{sec:analytic_bare_suscept} gives insight into the competition between ferromagnetic and altermagnetic instabilities and the mechanisms stabilizing these phases.

The leading altermagnetism implied from the above susceptibility analysis can be verified through selfconsistent Hartree-Fock calculations, an established method to examine instabilities in Hubbard models and recently discussed in view of two dimensional models for altermagnetism and antiferromagnetism~\cite{Maier2023}. Here, we start from the same tight-binding Hamiltonian and the Hubbard interaction, Eq.~(\ref{eq_hubbard}), perform a mean-field decoupling $n_{i,\mu,\sigma} \rightarrow \langle n_{i,\mu,\sigma} \rangle +\delta n_{i,\mu,\sigma}$, keep only terms quadratic in the fermionic operators and add the mean-field Hamiltonian $H_{\rm MF}=U\sum_{i,\mu,\sigma} \langle n_{i,\mu,\sigma} \rangle n_{i,\mu,\bar \sigma}$ to $H'$. The expectation value of the local density operator $\langle n_{i,\mu,\sigma}\rangle$ is then calculated using the eigenstates and eigenvalues in the symmetry-broken phase at a given temperature $T$. Iterations by updating the mean fields and the chemical potential while fixing the total number of electrons are performed until selfconsistency is achieved. In Fig.~\ref{fig:HF_RuO2}(e) we show the magnetic order parameter $|M|=\sum_{\mu}|\langle n_{\mu,\uparrow}\rangle-\langle n_{\mu,\downarrow}\rangle|$ for RuO$_2$ versus interaction strength $U$ for both altermagnetic and ferromagnetic order for a range of fillings $n$ indicated by the inset. As seen throughout the parameter regime, the altermagnetic instability dominates by exhibiting the smallest critical interaction strength.

\section{Berry curvature and crystal Hall effect}\label{sec:HallEffect}
In this section, we study altermagnetism in the presence of SOC, and derive a general analytic expression for the Berry curvature in the four-band case using the one-orbital minimal model introduced in Eq.~\eqref{eq:minimal_general_model}. Previous works have focused on finding an effective two-band Hamiltonian \cite{Smejkal2022Dec,Fang2023}, or solving numerically for the Berry curvature in the four bands case \cite{Smejkal2020Jun}. In general, these approaches give rise to a vanishing Hall conductivity quadratic in the SOC strength \cite{Smejkal2022Jun,Fang2023}, which is expected to be weak in altermagnets \cite{Smejkal2022Sep}. 
Here, we obtain an analytic expression for the Berry curvature linear in the SOC, leading to a larger non-vanishing Hall conductivity. To capture this result, care needs to taken in writing down the SOC term, for example replacing $\tau_y$ by $\tau_x$ in the SOC term \cite{Smejkal2020Jun,Fang2023} will yield a Berry curvature that is quadratic in SOC. 

\subsection{Berry curvature for the general minimal model}

Reference~\onlinecite{Graf2021Aug} describes an approach to obtain the Berry curvature for an N-band system without computing the eigenstates. In particular, they provide a general expression for the quantum geometric tensor
\begin{equation}
T_{n,ij}=\Tr{(\partial_i P_n)(1-P_n)(\partial_j P_{n})},
\label{eq:QMT_generaleq}
\end{equation}
where $P_n = \ket{\psi_n}\bra{\psi_n}$ is the projection operator onto band $n$. The Berry curvature corresponds to the imaginary part, $\Omega_{n,ij} = -2 \Im{T_{n,ij}}$, which is anti-symmetric in the indices $i,j$.

Without loss of generality, we consider SOC $\vec{\lambda}_{\kv}$ to have an arbitrary orientation and fix the altermagnetic  moments in-plane $\Vec{J}=(J,0,0)$. This moment direction choice is convenient for application to previous observations in altermagnetic material candidates, such as RuO$_2$~\cite{Feng2022Nov,Betancourt2023}. In this case, following Ref.~\cite{Graf2021Aug} and labeling the bands by $(\alpha,\beta)$ as in Eq.~\eqref{eq:dispersion_minimal_model}, we find the projection operator can be written as
\begin{equation}
    P_{\alpha,\beta} =  \frac{1}{4}\Big[ \mathds{1} + \frac{\widetilde{H}}{\widetilde{E}_\beta} \Big] \Big[ \mathds{1} + \frac{{H}}{{E}_{\alpha,\beta}} \Big],
\end{equation}
where $\widetilde{H} = t_{z,\kv} \sigma_x + \lambda_{y,\kv} \tau_x \sigma_z - \lambda_{z,\kv} \tau_x \sigma_y$ and $\widetilde{E}_{\beta=\pm} =  \beta \sqrt{t_{z,\kv}^2 + \lambda_{z,\kv}^2 + \lambda_{y,\kv}^2}$ are the corresponding eigenvalues. 

We find a Berry curvature linear in the SOC when the SOC term has the same spin direction as the altermagnetic order parameter $J$. In this case, to linear order, the Berry curvature becomes
\begin{equation}
    \begin{aligned}
        \Omega_{\alpha, \beta, ij} = & \frac{1}{E_{\alpha,\beta}^3} \sum_{m,n = i,j} \varepsilon_{mn} \Big[(J+\beta t_{z,\kv}) \partial_m \lambda_{x,\kv} \partial_n t_{x,\kv}  \\
        & + \beta\, t_{x,\kv} \partial_m t_{z,\kv} \partial_n \lambda_{x,\kv} + \beta\, \lambda_{x,\kv} \partial_m t_{x,\kv} \partial_n t_{z,\kv}\Big],
        \label{eq:general_Berrycurv_analytic}
    \end{aligned}
\end{equation}
with $\varepsilon_{mn}$ the antisymmetric tensor. This expression is relevant for any system with the four band Hamiltonian in Eq.~\eqref{eq:minimal_general_model}. Therefore, it is applicable to monoclinic, orthorhombic, tetragonal, rhombohedral, hexagonal, and cubic systems and the generality of this expression is one of the key result of this manuscript. This expression reveals that this Berry curvature result is {\it linear} in the component of the SOC {\it parallel} to the N\'eel order (we note that in RuO$_2$, we have found that the SOC we consider here is linear in the atomic SOC). This is very unlike the altermagnetic Rashba model \cite{Smejkal2022Dec}, where the Berry curvature is {\it quadratic} in the Rashba SOC and the relevant SOC is {\it perpendicular} to the N\'eel moment.

\subsection{Application to \texorpdfstring{RuO$_2$}{RuO2}}

For the tetragonal SG 136, i.e. RuO$_2$ and MnF$_2$,  the altermagnetic order parameter $\tau_z \sigma_z$ preserves the mirror symmetries $M_x, M_y, M_z$ and therefore there is a vanishing anomalous Hall effect and Berry phase. Focusing on the tetragonal case, $\tau_z \sigma_x$ and $\tau_z \sigma_y$ can generate an anomalous Hall effect in the presence of SOC. In particular, the former term breaks the $M_x, M_z$ mirrors and, as a consequence, it can generate a finite Hall conductivity $\sigma_{xz}$, which is given by the integral over the filled bands of the Berry curvature
\begin{equation}
    \sigma_{ij} = - \frac{e^2}{\hbar} \int_{\textrm{BZ}} \frac{d\kv}{(2\pi)^3} \sum_{n} f_n(\kv) \Omega_{n,ij},
    \label{eq:conductivity}
\end{equation}
with $f_n(\kv)$ the Fermi-Dirac distribution of each band $n$.

Recalling the form of the coefficients in the tight-binding Hamiltonian given in Eqs.~\eqref{eq:xyorb_SOC},\eqref{eq:xyorb_hoppings} for tetragonal systems with the rutile structure, the Berry curvature in Eq.~\eqref{eq:general_Berrycurv_analytic} gives
\begin{equation}
    \Omega_{\alpha,\beta, xz} = \frac{1}{8 E_{\alpha,\beta}^3} \lambda  t_8 J \cos^2\left(\frac{k_y}{2}\right) (\cos k_z-\cos k_x),
    \label{eq:Berrycurv_analytic4x4}
\end{equation}
including only the terms that give rise to a non-vanishing Hall conductivity, i.e., do not average to zero in the BZ integral of Eq.~\eqref{eq:conductivity}.

\begin{figure}[t]
\begin{center}
\includegraphics[angle=0,width=.95\linewidth]{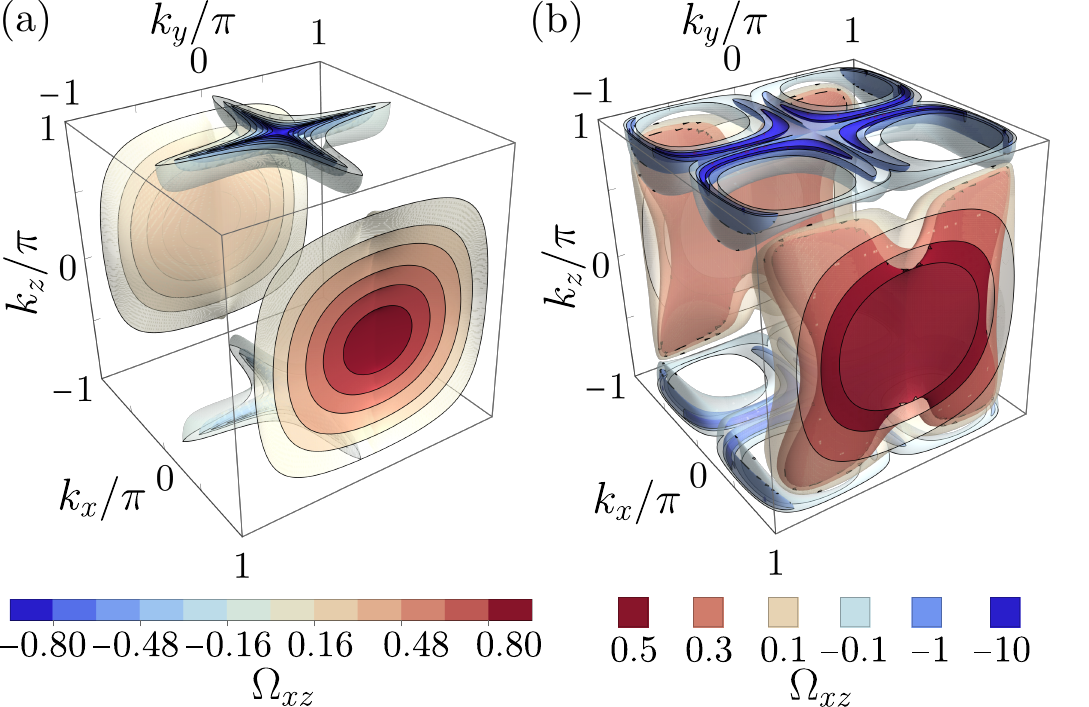}
\caption{Three-dimensional Berry curvature obtained from the analytic expression in Eq.~\eqref{eq:Berrycurv_analytic4x4}, with $\alpha=+$ and (a) $\beta=+$ and (b) $\beta=-$, considering the one-orbital minimal model [Eqs.~\eqref{eq:xyorb_hoppings},\eqref{eq:xyorb_SOC}] normal state band structure for RuO$_2$ shown in Fig.~\ref{fig:RuO2_bands}(a). We choose $J=0.2$ for the altermagnetic order parameter and estimate the SOC $\lambda = 0.1$ from the relativistic DFT calculations shown in Fig.~\ref{fig:relativisticDFT}.}
\label{fig:Berry_curv}
\end{center}
\end{figure}

In Fig.~\ref{fig:Berry_curv} we show the Berry curvature for the $\alpha = + $ and $\beta = \pm$ bands, considering the realistic hopping parameters in Fig.~\ref{fig:RuO2_bands}(a) for the one-orbital model reproducing the RuO$_2$ band structure. The same plots are obtained for $\alpha = -$ by exchanging the red and blue colors due to inversion symmetry. 
We have estimated a realistic SOC strength in RuO$_2$ from the splitting in the Z-R-A path in the relativistic DFT band structure, as shown in Fig.~\ref{fig:relativisticDFT} in Appendix \ref{ap:orbital_projection}. In agreement with previous DFT calculations, the SOC has a weak effect on the bands \cite{Smejkal2022Sep}.
Figure~\ref{fig:Berry_curv}(a) shows that $\Omega_{xz}$ is large at the nodal planes X-M and Z-R, where the normal state band structure shown in Fig.~\ref{fig:RuO2_bands}(a) features degeneracies. As discussed in Sec.~\ref{sec:susceptibilities}, the band degeneracies lead to a large interband susceptibility, thus favoring altermagnetism. Consequently, in the presence of an altermagnetic instability the Berry curvature is guaranteed to be large at the nodal planes.

\begin{figure}[t]
\begin{center}
\includegraphics[width=\linewidth]{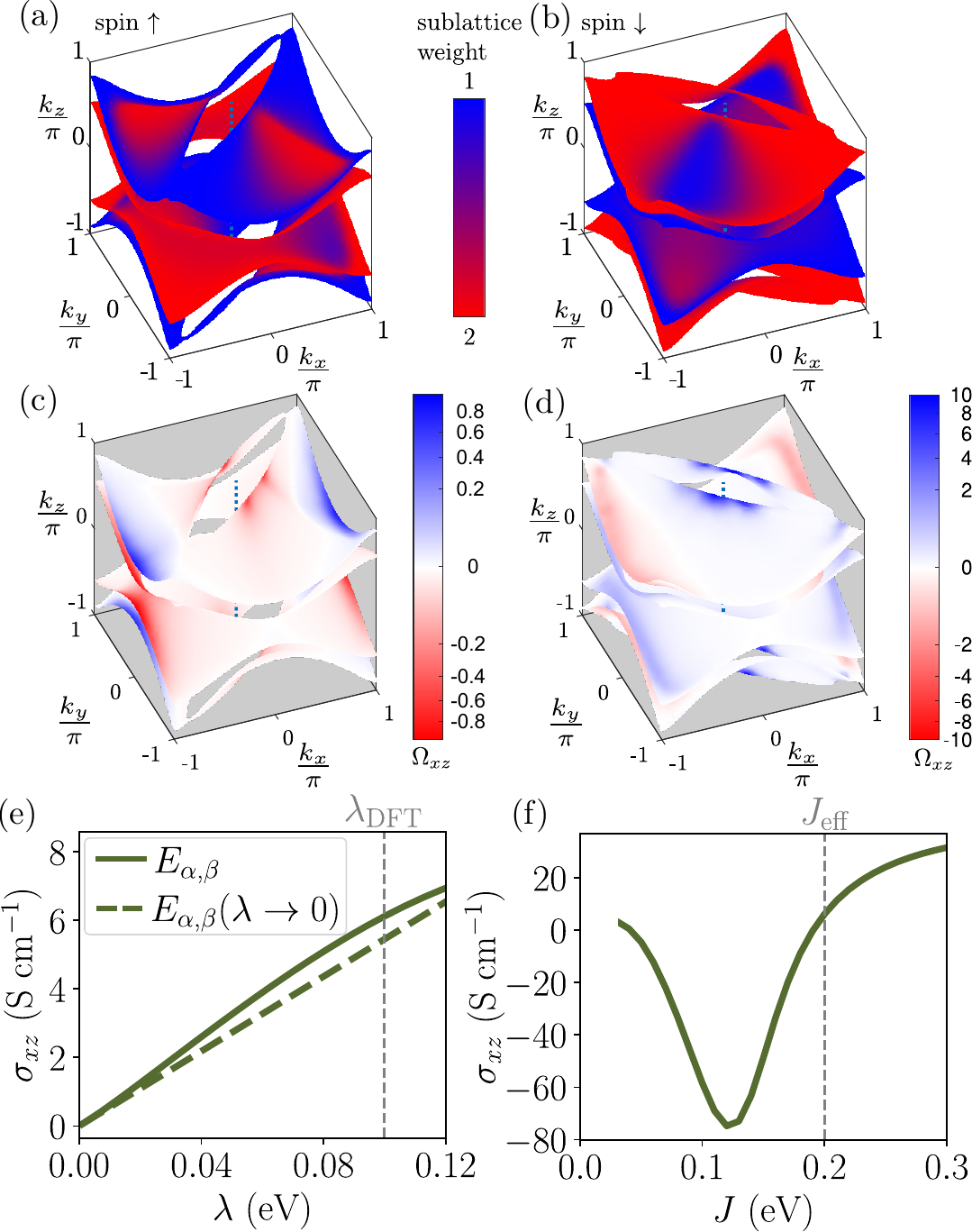}
\caption{(a)-(b) Spin-up and spin-down Fermi surfaces for RuO$_2$ in the altermagnetic state, considering the minimal model in Eq.~\eqref{eq:minimal_general_model} and the state band structure shown in Fig.~\ref{fig:RuO2_bands}. The colorbar indicates sublattice weight. (c)-(d) Berry curvature shown in Fig.~\ref{fig:Berry_curv} projected onto the spin-up and spin-down Fermi surfaces. 
(e)-(f) Conductivity as a function of SOC strength, with $J=0.2$ and $n_\kv = 201^3$, and magnetic order, with $\lambda=0.1$ and $n_\kv = 401^3$, for $T=0.01$. $\lambda_{\textrm{DFT}}$ and $J_{\rm{eff}}$ denote the effective SOC and magnetic moment obtained by comparing with DFT results (see Appendix~\ref{ap:orbital_projection} and Ref.~\cite{Ahn2019}).
}
\label{fig:RuO2_conductivity}
\end{center}
\end{figure}

In addition, as previously discussed in Sec.~\ref{sec:minimal_TBH}, Fig.~\ref{fig:Berry_curv}(b) shows that the presence of Weyl loops further enhances the Berry curvature~\cite{Smejkal2022Jun}. Neglecting the SOC terms, the eigenenergies for the minimal model in Eq.~\eqref{eq:dispersion_minimal_model} with in-plane moment $J$ correspond to $E_{\alpha = \pm,\beta = \pm} = \alpha \sqrt{t_{x,\kv}^2 + (J+\beta t_{z,\kv})^2}$. When $t_{x,\kv}=0$, one band is two-fold degenerate for $t_{z,\kv} = \pm J$. The band degeneracies stem from Weyl loops on the $k_z = \pi$ face, which manifest as band crossings along the Z-A direction in Fig.~\ref{fig:RuO2_bands}.

Remarkably, Fig.~\ref{fig:Berry_curv} can be used as an indication of the most favorable regions that the Fermi surface should touch in order to obtain a large Berry curvature, thus giving rise to a large anomalous Hall response. As already suggested for FeSb$_2$ in Ref.~\onlinecite{Mazin2021}, the chemical doping can be used the push the Fermi surface to a large curvature $\Omega$ region.

In Figs.~\ref{fig:RuO2_conductivity}(a) and \ref{fig:RuO2_conductivity}(b) we show the spin-up and spin-down Fermi surfaces for RuO$_2$, respectively, indicating the sublattice weight. Panels (c) and (d) of Fig.~\ref{fig:RuO2_conductivity} display the projected Berry curvature from Fig.~\ref{fig:Berry_curv} onto the two Fermi surfaces. As seen, when summing the contributions for the two Fermi surfaces this will give rise to a non-vanishing anomalous crystal Hall effect. Figure~\ref{fig:RuO2_conductivity}(e) explicitly shows the conductivity calculated using Eq.~\eqref{eq:conductivity}. In the limit $\lambda\rightarrow 0 $, the conductivity scales linearly with SOC (dashed line) and this parameter is directly proportional to the atomic SOC~\cite{Clepkens2021}. Sub-leading non-linear contributions are also present due to the non-trivial dependence of $E_{\alpha,\beta}$ on the SOC (solid line), as seen from Eq.~\eqref{eq:dispersion_minimal_model}. For stoichiometric RuO$_2$, DFT results predicted $\sigma_{xz} = 36.4$ S cm$^{-1}$ \cite{Smejkal2020Jun}. Therefore, considering the linear in SOC contribution we obtain a large conductivity. Note that the calculation has been done using the minimal one-orbital model in Eq.~\eqref{eq:minimal_general_model} and, as a consequence, we expect contributions from other bands in RuO$_2$, see Fig.~\ref{fig:RuO2_bands}(a). Figure~\ref{fig:RuO2_conductivity}(f) displays the conductivity $\sigma_{xz}$ as a function of magnetic order $J$. As seen, $\sigma_{xz}$ exhibits a significant $J$ dependence both in terms of amplitude and sign. A similar strong dependence of the anomalous Hall response to band structure details has been recently discussed for FeSb$_2$ in Ref.~\onlinecite{Mazin2021}. These results suggest that one might significantly enhance the Hall conductivity by band engineering or optimization of the altermagnetic order parameter.

\section{Minimal models for other altermagnetic candidates}\label{sec:TBM_AMcandidates}
In this section, we demonstrate that the general minimal model in Eq.~\eqref{eq:minimal_general_model} can also be used to describe other altermagnetic candidates with different symmetry properties~\cite{Smejkal2022Dec,Smejkal2022Sep}. Here we initially present results on the stability of $d$-wave altermagnetism for orthorhombic FeSb$_2$. We then develop minimal models for $g$-wave altermagnetism in hexagonal materials such as CrSb and MnTe and for $i$-wave altermagnetism in cubic materials. In Appendix~\ref{ap:organic} we use the minimal model to describe the bands for the organic compound $\kappa$-Cl, which provides a platform to study 2D altermagnetism. Motivated by this, Appendix~\ref{ap:2D_tetragonal_toymodel} presents a minimal 2D model for altermagnetism in a tetragonal system.

\subsection{\texorpdfstring{FeSb$_2$}{FeSb2}}
\begin{figure}[t!]
\begin{center}
\includegraphics[angle=0,width=.95\linewidth]{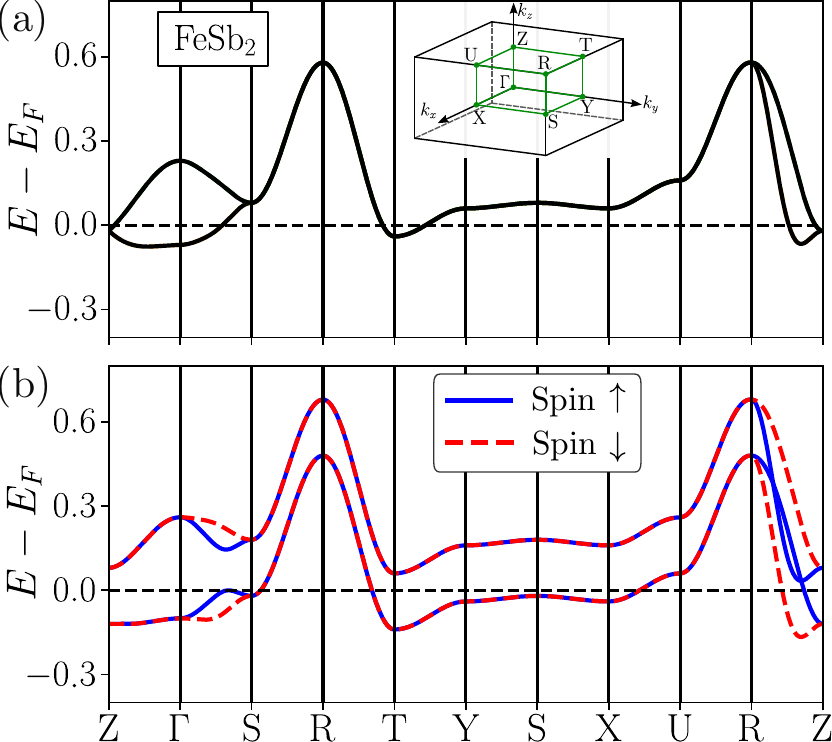}
\caption{Normal state (a) and altermagnetic (b) band structures inspired by FeSb$_2$ obtained from the minimal model in Eq.~\eqref{eq:minimal_general_model} considering Eqs.~\eqref{eq:xyorb_hoppings},\eqref{eq:disperion_FeSb2}, with hopping parameters detailed in Appendix~\ref{ap:hoppings} and $J_z = 0.1$ in (b). The BZ path is shown in the inset.}
\label{fig:bands_FeSb2}
\end{center}
\end{figure}

Initially FeSb$_2$ was proposed to be a ferromagnet \cite{Lukoyanov2006}, although Ref.~\onlinecite{Mazin2021} recently suggested that this material is nonmagnetic and, more intriguingly, the doped compound could host unconventional magnetism. 
FeSb$_2$ is an orthorhombic material with space group 58. This space group has point group $D_{2h}$ and Fe occupies the 2a Wyckoff position with site symmetry $C_{2h}$, see row 10 in Table \ref{tab:spacegroups_mo}. This implies that the $\tau_z$ and $\tau_y$ operators belong to the $B_{1g}$ representation of $D_{2h}$. Based on Ref.~\onlinecite{Lukoyanov2006}, we construct a minimal model for $d_{x^2-y^2}$-orbitals, considering  Wyckoff positions at $(0,0,0)$ and $(1/2,1/2,1/2)$. Thus, the minimal model in Eq.~\eqref{eq:minimal_general_model} in this case has the same form for the hoppings $t_{x,\kv}$ and $t_{z,\kv}$ as in Eq.~\eqref{eq:xyorb_hoppings}. The SOC is given by
\begin{equation}
    \begin{aligned}
        & \lambda_{x,\kv} = \lambda_x \sin \frac{k_z}{2} \sin \frac{k_x}{2} \cos \frac{k_y}{2}, \\
        & \lambda_{y,\kv} = \lambda_y \sin \frac{k_z}{2} \sin \frac{k_y}{2} \cos \frac{k_x}{2}, \\
        & \lambda_{z,\kv} = \lambda_z \cos \frac{k_z}{2} \cos \frac{k_x}{2} \cos \frac{k_y}{2},
    \end{aligned}
    \label{eq:xyorb_SOC_2}
\end{equation}
and the dispersion $\varepsilon_{0,\kv}$ corresponds to
\begin{align}
{\varepsilon_{0,\kv}} &=t_{1x} \cos k_x + t_{1y} \cos k_y + t_2 \cos k_z + t_3 \cos k_x \cos k_y\notag \\
        &\phantom{=}+ t_{4x} \cos k_x \cos k_z + t_{4y} \cos k_y \cos k_z \notag \\
        &\phantom{=}+ t_5 \cos k_x \cos k_y \cos k_z - \mu.
        \label{eq:disperion_FeSb2}
\end{align}
Figure~\ref{fig:bands_FeSb2}(a) displays a single band picture inspired by the FeSb$_2$ band structure that crosses the Fermi energy at the R-point \cite{Lukoyanov2006,Mazin2021}. Importantly, the spin splitting along the $\Gamma$-S and R-Z directions (see inset in Fig.~\ref{fig:bands_FeSb2}(a)) can also be described by the $t_{z,\kv} \tau_z$ term, as seen in Fig.~\ref{fig:bands_FeSb2}(b).

In agreement with Ref.~\onlinecite{Mazin2021}, only when lowering the chemical potential we obtain a leading altermagnetic instability, see also Appendix~\ref{ap:HF_organic_FeSb2}.
In contrast to the previous results for other compounds, in this case an RPA analysis reveals that the altermagnetic susceptibility does not diverge at $\qv \rightarrow 0$, pointing to an incommensurate altermagnetic state as the leading instability due to Fermi surface nesting (see Appendix~\ref{ap:HF_organic_FeSb2}). However, Ref.~\onlinecite{Mazin2021} only considered a few commensurate vectors, whereas our RPA analysis shows a different $\qv \neq 0$ leading divergence.

Notably, in FeSb$_2$ there is a crucial difference compared to the other materials discussed in this work. As seen from Fig.~\ref{fig:bands_FeSb2}(a), along the R-Z line the two bands disperse downwards. Therefore, we expect that altermagnetism in this compound is not as stable as in the case of RuO$_2$ and $\kappa$-Cl (see Appendix~\ref{ap:organic}), where the bands close to the A and the S-point, respectively, disperse in opposite directions giving rise to a van Hove singularity. As shown in Appendix \ref{ap:HF_organic_FeSb2}, upon small variations of the filling, altermagnetism becomes unstable and ferromagnetism is the leading instability. Finally, we would like to point out that these predictions should be revisited when considering a more complete model including the full set of $d$-orbitals. Nevertheless, the single-orbital model describes the right spin splitting and allows us to derive an expression for the Berry curvature, written in Eq.~\eqref{eq:Berrycurv_analytic4x4}, which is large at the nodal planes and enhanced by Weyl lines, giving rise to a non-vanishing conductivity.

\subsection{\texorpdfstring{$g$}{g}-wave altermagnetism}

The case of CrSb is interesting since a particularly large band splitting as high as 1.2 eV has been predicted for this compound \cite{Smejkal2022Sep,Smejkal2022Dec}, and the altermagnetic band splitting has been recently observed experimentally \cite{Reimers2024}.
CrSb is a metal with an hexagonal structure and a critical temperature of 705~K, thus already magnetically ordered at room temperature \cite{Park2020Dec}. Motivated by this material and by MnTe \cite{Lee2024}, we give a minimal model that gives rise to $g$-wave altermagnetism. Both 
CrSb and MnTe belong to the space group 194, with point group $D_{6h}$. The Cr and Mn atoms form an hexagonal lattice with Wyckoff position 2a, which has site symmetry $D_{3d}$. Consequently, the $\tau_z$ and $\tau_y$ operators belong to the $B_{1g}$ representation of $D_{6h}$, see Table \ref{tab:spacegroups_trhc}. Incorporating $\tv=(0,0,1/2)$ the minimal model in Eq.~\eqref{eq:minimal_general_model} has the form

\begin{equation}
    \varepsilon_{0,\kv} = t_1 \left(\cos k_x + 2 \cos \frac{k_x}{2} \cos\frac{\sqrt{3}k_y}{2} \right) + t_2 \cos k_z - \mu,
\end{equation}
while for the hoppings with $\tau_z$ and $\tau_x$ we obtain
\begin{equation}
    \begin{aligned}
        & t_{x,\kv} = t_3 \cos\frac{k_z}{2},\\
        & t_{z,\kv} = t_4 \sin k_z f_y (f_y^2-3f_x^2), 
    \end{aligned}
\end{equation}
and  SOC 
\begin{equation}
    \begin{aligned}
        & \lambda_{x,\kv} = \lambda \cos\frac{k_z}{2}(f_x^2 - f_y^2),\\
        & \lambda_{y,\kv} = - 2\lambda \cos\frac{k_z}{2} f_x f_y, \\
        & \lambda_{z,\kv} = \lambda_z \sin\frac{k_z}{2}f_x (f_x^2 - 3f_y^2),
    \end{aligned}    
\end{equation}
where we have defined $f_x \equiv  \sin k_x+\sin\frac{k_x}{2}\cos\frac{\sqrt{3}k_y}{2}$ and $f_y\equiv \sqrt{3}\cos\frac{k_x}{2}\sin\frac{\sqrt{3}k_y}{2}$, see also Table \ref{tab:models_nodal}.

In agreement with the orthorhombic and tetragonal space groups, for the hexagonal case the symmetry of the term coupling to $\tau_z$ gives the symmetry of the altermagnetic splitting, which corresponds to a $g$-wave. Therefore, also in this case the one-orbital minimal model in Eq.~\eqref{eq:minimal_general_model} is sufficient to describe the symmetry of the spin splitting. However, in contrast to the examples of RuO$_2$ and FeSb$_2$ where only non-degenerate 1D orbital IRs of $S$ exist, the case $S=D_{3d}$ also allows 2D orbital IRs. In this case, it would be of interest to extend our microscopic model to include these 2D degenerate orbital IRs since both degenerate 2D IRs and non-degenerate 1D IRs are relevant to the electronic structure of  CrSb and MnTe.

\subsection{\texorpdfstring{$i$}{i}-wave altermagnetism}
In the previous cases we have shown that the minimal model in Eq.~\eqref{eq:minimal_general_model} can give rise to $d$-wave and $g$-wave altermagnetism. Here, we demonstrate that it is also sufficiently general to allow for $i$-wave altermagnetism. In particular, we focus on the cubic SG 223 with point group $O_h$ and  Wyckoff position 2a with site symmetry $T_h$. Consequently, the $\tau_y$ and $\tau_z$ operators belong to the $A_{2g}$ representation of the point group $O_h$, see Table \ref{tab:spacegroups_trhc}. The translation between the two magnetic atoms in the unit cell is $\tv = (\frac{1}{2},\frac{1}{2},\frac{1}{2})$. Therefore, the minimal model in Eq.~\eqref{eq:minimal_general_model} has the following form (Table \ref{tab:models_no_nodal}):
\begin{equation}
\begin{aligned}
    &\varepsilon_{0,\kv} = t_1 (\cos k_x + \cos k_y + \cos k_z) - \mu, \\
    &t_{x,\kv} = t_2 \cos \frac{k_x}{2} \cos\frac{k_y}{2} \cos\frac{k_z}{2}, \\
    &t_{z,\kv} = t_3 (\cos k_x {-} \cos k_y) (\cos k_x {-} \cos k_z) (\cos k_y {-} \cos k_z),
\end{aligned}
\end{equation}
with SOC given by
\begin{equation}
\begin{aligned}
    &\lambda_{x,\kv} = \lambda \sin \frac{k_y}{2} \sin\frac{k_z}{2}\cos\frac{k_x}{2}, \\
    &\lambda_{y,\kv} = \lambda \sin\frac{k_x}{2}\sin\frac{k_z}{2}\cos\frac{k_y}{2}, \\
    &\lambda_{z,\kv} = \lambda \sin\frac{k_x}{2}\sin\frac{k_y}{2}\cos\frac{k_z}{2}.
\end{aligned}  
\end{equation}
Since the altermagnetic spin-splitting is given by $t_{z,\kv}$ and since, 
in this case, $t_{z,\kv}$ belongs to the $A_{2g}$ representation of $O_h$ and near the $\Gamma$ point, $t_{z,\kv}\sim x^4(y^2-z^2) + y^4(z^2-x^2)+z^4(x^2-y^2)$, this model describes an $i$-wave altermagnet.

\section{Discussion and conclusions}\label{sec:discussion}

We have provided minimal models for orthorhombic, tetragonal, hexagonal, and cubic space groups. While we did not give examples, our approach can also be applied to monoclinic and rhombohedral space groups, see Tables \ref{tab:spacegroups_mo} and \ref{tab:spacegroups_trhc}. For the monoclinic case, an example is SG 14, with point group $C_{2h}$, Wyckoff position 2a or 2b (this applies to the Re site in K$_2$ReI$_6$ \cite{Guo2023Mar}), and site symmetry $C_i$, our approach will give a corresponding  minimal model for a $d$-wave altermagnet. For the rhombohedral case, an example is SG 167, with point group $D_{3d}$, Wyckoff position 6b (this applies to the Fe site in FeCo$_3$ \cite{Guo2023Mar}), and site symmetry $S_6$, our approach will give a corresponding minimal model for a $g$-wave altermagnet.

Our minimal models also highlight the role of non-symmorphic band degeneracies in altermagnetism. In particular,  the altermagnetic susceptibility we introduce reveals that band degeneracies help to stabilize the altermagnetic state. This susceptibility further reveals that the presence of van Hove singularities related to the band degeneracies are favorable for altermagnetism. Indeed, in the case of FeSb$_2$, we find that the corresponding lack of a van Hove singularity may lead to an incommensurate altermagnetic state. In addition, the non-symmorphic band degeneracies enhance the Berry curvature and thus lead to a large crystal Hall effect. 

The minimal models can also be used to obtain new insight into the properties of altermagnets. For example, as outlined in Appendix \ref{ap:topological_states}, we show how our minimal model can give rise to the existence of topological states and  chiral surface bound states. In addition, superconductivity has been observed in strained RuO$_2$ \cite{Uchida2020,Ruf2021,Occhialini2022}, which is surprising since altermagnetism tends to strongly suppress superconductivity. Our minimal model suggests an answer for this: in the presence of strain $\epsilon_{xy}$, the term 
$\epsilon_{xy}  \cos k_x \cos k_y \tau_z$ also appears in the Hamiltonian. This term splits the van Hove singularity at the A-point. As a consequence, we expect that altermagnetism will be suppressed, potentially favoring other nearby electronic instabilities. We expect that our minimal models will serve as a useful tool to examine spatial varying properties of altermagnets, such as magnetic domain walls, and will shed insight into the interplay of other electronic instabilities, such as superconductivity, with altermagnetism. 

In conclusion, through the comparison to DFT results, we have developed realistic models for altermagnetism based on a two magnetic atom sublattice in non-symmorphic materials. These models can be applied to monoclinic, orthorhombic, tetragonal, rhombohedral, and cubic point crystals and can describe $d$-wave, $g$-wave, and $i$-wave altermagnets. Furthermore, we have shown that these models generically give rise to a Berry curvature that is linear in the spin-orbit coupling.  We expect that these minimal models will serve as a useful tool to understand altermagnetism and its properties.

\section*{Acknowledgments} 
We thank L. Li, D. Radevych, T. Shishidou,  S. Sumita, and M. Weinert for useful discussions. 
M.~R. acknow\-ledges support from the Novo Nordisk Foundation grant NNF20OC0060019. A.~K. acknowledges support by the Danish National Committee for Research Infrastructure (NUFI) through the ESS-Lighthouse Q-MAT. D.~F.~A. and Y.~Y. were supported by the National Science Foundation Grant No. DMREF 2323857.

\appendix
\renewcommand{\thefigure}{S\arabic{figure}}
\setcounter{figure}{0}

\begin{widetext}
	\clearpage 
\section{Tables of minimal models}\label{ap:otherspacegroups}

Here we provide explicit minimal models for all space groups in Tables \ref{tab:spacegroups_mo} and \ref{tab:spacegroups_trhc} that have primitive lattice structures. We provide two Tables. Table \ref{tab:models_nodal} contains only non-symmorphic space groups with nodal planes in the paramagnetic state. These nodal planes have four-fold degenerate fermions that exist when spin-orbit coupling is not included, and they appear on a planar face of the Brillouin zone. Table \ref{tab:models_no_nodal} contains minimal models for symmorphic groups and all non-symmorphic groups that do not contain nodal planes.
\begin{table}[h]
\begin{tabular}{|c|c|c|c|c|c|}
\hline SG&$\tau_x$&$\tau_z$&$\tau_y\sigma_x$&$\tau_y\sigma_y$&$\tau_y\sigma_z$ \\ \hline
11(2a-2d)&$c_{y/2}$&$s_y(s_x,s_z)$&$c_{y/2}$&$s_{y/2}(s_x,s_z)$&$c_{y/2}$\\ \hline
14(2a-2d)&$c_{y/2}(s_xs_{z/2},c_{z/2})$&$s_y(s_x,s_z)$&$c_{y/2}(s_xs_{z/2},c_{z/2})$&$s_{y/2}(s_xc_{z/2},s_{z/2})$&$c_{y/2}(s_xs_{z/2},c_{z/2})$\\ \hline
51(2a-2d)&$c_{x/2}$&$s_xs_z$&$s_{x/2}s_y$&$c_{x/2}$&$c_{x/2}s_ys_z$\\ \hline
53(2a-2d)&$c_{x/2}c_{z/2}$&$s_ys_z$&$c_{x/2}c_{z/2}$&$s_{x/2}s_yc_{z/2}$&$s_{x/2}s_{z/2}$\\ \hline
55(2a-2d)&$c_{x/2}c_{y/2}$&$s_xs_y$&$s_{x/2}c_{y/2}s_z$&$c_{x/2}s_{y/2}s_z$&$c_{x/2}c_{y/2}$\\ \hline
58(2a-2d)&$c_{x/2}c_{y/2}c_{z/2}$&$s_xs_y$&$s_{x/2}c_{y/2}s_{z/2}$&$c_{x/2}s_{y/2}s_{z/2}$&$c_{x/2}c_{y/2}c_{z/2}$ \\ \hline
127(2a,2b)&$c_{x/2}c_{y/2}$&$s_xs_y(c_{x}-c_{y})$&$\lambda_{}s_{x/2}c_{y/2}s_z$&$\lambda_{}c_{x/2}s_{y/2}s_z$&$c_{x/2}c_{y/2}$\\ \hline
127(2c,2d)&$c_{x/2}c_{y/2}$&$s_xs_y$&$\lambda_{}s_{x/2}c_{y/2}s_z$&$-\lambda_{}c_{x/2}s_{y/2}s_z$&$c_{x/2}c_{y/2}(c_{x}-c_{y})$\\ \hline
128(2a,2b)&$c_{x/2}c_{y/2}c_{z/2}$&$s_xs_y(c_{x}-c_{y})$&$\lambda_{}s_{x/2}c_{y/2}s_{z/2}$&$\lambda_{}c_{x/2}s_{y/2}s_{z/2}$&$c_{x/2}c_{y/2}c_{z/2}$ \\ \hline
136(2a,2b)&$c_{x/2}c_{y/2}c_{z/2}$&$s_xs_y$&$\lambda_{}s_{x/2}c_{y/2}s_{z/2}$&$-\lambda_{}c_{x/2}s_{y/2}s_{z/2}$&$c_{x/2}c_{y/2}c_{z/2}(c_{x}-c_{y})$ \\ \hline
176(2b)&$c_{z/2}$&$\begin{matrix}s_{z}f_{x}(f_{x}^2-3f_{y}^2),\\s_{z}f_{y}(f_{y}^2-3f_{x}^2)\end{matrix}$&$\begin{matrix}\lambda_1c_{z/2}(f_x^2-f_y^2)\\+2\lambda_2c_{z/2}f_xf_y\end{matrix}$ &$\begin{matrix}-2\lambda_1c_{z/2}f_xf_y\\+\lambda_2c_{z/2}(f_x^2-f_y^2)\end{matrix}$ &$\begin{matrix}s_{z/2}f_{x}(f_{x}^2-3f_{y}^2),\\s_{z/2}f_{y}(f_{y}^2-3f_{x}^2)\end{matrix}$ \\ \hline
193(2b)&$c_{z/2}$&$s_{z}f_{x}(f_{x}^2-3f_{y}^2)$&$2\lambda_{}c_{z/2}f_xf_y$ &$\lambda_{}c_{z/2}(f_x^2-f_y^2)$ &$s_{z/2}f_{y}(f_{y}^2-3f_{x}^2)$ \\ \hline
194(2a)&$c_{z/2}$&$s_{z}f_{y}(f_{y}^2-3f_{x}^2)$&$\lambda_{}c_{z/2}(f_x^2-f_y^2)$ &$-2\lambda_{}c_{z/2}f_xf_y$ &$s_{z/2}f_{x}(f_{x}^2-3f_{y}^2)$ \\ \hline
\end{tabular}
\caption{Tight-binding coefficients for space groups with a nodal plane and two atoms per unit cell at the inversion center. Abbreviation $c_i\equiv\cos k_i$, $s_i\equiv\sin k_i$, $c_{i/2}\equiv\cos \frac{k_i}{2}$, $s_{i/2}\equiv\sin \frac{k_i}{2}$,$f_x\equiv\sin k_x+\sin\frac{k_x}{2}\cos\frac{\sqrt{3}k_y}{2}$, and $f_y\equiv \sqrt{3}\cos\frac{k_x}{2}\sin\frac{\sqrt{3}k_y}{2}$ applies. We note that while products of $f_x$ and $f_y$ are convenient for describing the symmetry of the appropriate terms in the Hamiltonian, these products generally contain nearest neighbor hopping terms together with longer range hopping terms. Within tetragonal and hexagonal space groups, the coefficients for the two SOC $\tau_y\sigma_{x,y}$ are related. The other coefficients, which are generally different, are omitted from this table. For instance, $s_y(s_x,s_z)$ in SG11 means $t_{z1}s_ys_x+t_{z2}s_ys_z$.}\label{tab:models_nodal}
\end{table}

\begin{table}[h]
\begin{tabular}{|c|c|c|c|c|c|}
\hline SG&$\tau_x$&$\tau_z$&$\tau_y\sigma_x$&$\tau_y\sigma_y$&$\tau_y\sigma_z$ \\ \hline
13(2a-2d)&$c_{z/2},s_xs_{z/2}$&$s_y(s_x,s_z)$&$c_{z/2},s_xs_{z/2}$&$s_y(s_{z/2},s_xc_{z/2})$&$c_{z/2},s_xs_{z/2}$\\ \hline
49(2a-2d)&$c_{z/2}$&$s_xs_y$&$s_xs_{z/2}$&$s_ys_{z/2}$&$c_{z/2}$\\ \hline
83(2e,2f)&$\begin{matrix}c_{x/2}c_{y/2},\\s_{x/2}s_{y/2}(c_x-c_y)\end{matrix}$&$(c_x-c_y),s_xs_y$&$\begin{matrix}\lambda_1s_{x/2}c_{y/2}s_z\\+\lambda_2c_{x/2}s_{y/2}s_z\end{matrix}$&$\begin{matrix}-\lambda_1c_{x/2}s_{y/2}s_z\\+\lambda_2s_{x/2}c_{y/2}s_z\end{matrix}$&$\begin{matrix}s_{x/2}s_{y/2},\\c_{x/2}c_{y/2}(c_x-c_y)\end{matrix}$\\ \hline
84(2a,2b)&$c_{z/2}$&$(c_x-c_y),s_xs_y$&$(\lambda_1s_x+\lambda_2s_y)s_{z/2}$&$(-\lambda_1s_y+\lambda_2s_x)s_{z/2}$&$c_{z/2}(c_x-c_y)$\\ \hline
84(2c,2d)&$\begin{matrix}c_{x/2}c_{y/2}c_{z/2},\\s_{x/2}s_{y/2}c_{z/2}\end{matrix}
$&$(c_x-c_y),s_xs_y$&$\begin{pmatrix}+\lambda_1s_{x/2}c_{y/2}\\+\lambda_2c_{x/2}s_{y/2}\end{pmatrix}s_{z/2}$&$\begin{pmatrix}-\lambda_1c_{x/2}s_{y/2}\\+\lambda_2s_{x/2}c_{y/2}\end{pmatrix}s_{z/2}$&$\begin{matrix}c_{x/2}c_{y/2}c_{z/2}(c_x-c_y),\\s_{x/2}s_{y/2}c_{z/2}(c_x-c_y)\end{matrix}$\\ \hline
123(2e,2f)&$c_{x/2}c_{y/2}$&$(c_x-c_y)$&$\lambda_{}c_{x/2}s_{y/2}s_z$&$\lambda_{}s_{x/2}c_{y/2}s_z$&$s_{x/2}s_{y/2}$\\ \hline
124(2b,2d)&$c_{z/2}$&$s_xs_y(c_x-c_y)$&$\lambda_{}s_{x}s_{z/2}$&$\lambda_{}s_{y}s_{z/2}$&$c_{z/2}$\\ \hline
131(2a,2b)&$c_{z/2}$&$(c_x-c_y)$&$\lambda_{}s_{y}s_{z/2}$&$\lambda_{}s_{x}s_{z/2}$&$s_{x}s_{y}c_{z/2}$\\ \hline
131(2c,2d)&$c_{x/2}c_{y/2}c_{z/2}$&$(c_x-c_y)$&$\lambda_{}c_{x/2}s_{y/2}s_{z/2}$&$\lambda_{}s_{x/2}c_{y/2}s_{z/2}$&$s_{x/2}s_{y/2}c_{z/2}$\\ \hline
132(2a,2c)&$c_{z/2}$&$s_xs_y$&$\lambda_{}s_{x}s_{z/2}$&$-\lambda_{}s_{y}s_{z/2}$&$c_{z/2}(c_x-c_y)$\\ \hline
163(2b)&$\begin{matrix} c_{z/2},\\f_x(3f_y^2-f_x^2)s_{z/2}\end
{matrix}$&$\begin{matrix} f_y(f_y^2-3f_x^2)s_{z},\\f_xf_y(f_x^2-3f_y^2)(3f_x^2-f_y^2)\end
{matrix}$&$\begin{matrix} \lambda_1f_xs_{z/2}\\+\lambda_2(f_x^2-f_y^2)c_{z/2}\end
{matrix}$&$\begin{matrix} \lambda_1f_ys_{z/2}\\-2\lambda_2f_xf_yc_{z/2}\end
{matrix}$&$\begin{matrix} 
c_{z/2},\\f_x(3f_y^2-f_x^2)s_{z/2}\end
{matrix}$ \\ \hline
165(2b)&$\begin{matrix} c_{z/2},\\f_y(3f_x^2-f_y^2)s_{z/2}\end
{matrix}$&$\begin{matrix} f_x(f_x^2-3f_y^2)s_{z},\\f_xf_y(f_x^2-3f_y^2)(3f_x^2-f_y^2)\end
{matrix}$&$\begin{matrix} \lambda_1f_xs_{z/2}\\+2\lambda_2f_xf_yc_{z/2}\end
{matrix}$&$\begin{matrix} \lambda_1f_ys_{z/2}\\+\lambda_2(f_x^2-f_y^2)c_{z/2}\end
{matrix}$&$\begin{matrix} c_{z/2},\\f_y(3f_x^2-f_y^2)s_{z/2}\end
{matrix}$ \\ \hline
192(2b)&$c_{z/2}$&$f_xf_y(f_x^2-3f_y^2)(3f_x^2-f_y^2)$&$\lambda_{}s_{z/2}f_x$ &$\lambda_{}s_{z/2}f_y$&$c_{z/2}$ \\ \hline
223(2a)&$c_{x/2}c_{y/2}c_{z/2}$&$(c_x-c_y)(c_y-c_z)(c_z-c_x)$&$\lambda_{}c_{x/2}s_{y/2}s_{z/2}$&$\lambda_{}s_{x/2}c_{y/2}s_{z/2}$&$\lambda_{}s_{x/2}s_{y/2}c_{z/2}$\\ \hline
\end{tabular}
\caption{Tight-binding coefficients for space groups with two atoms per unit cell at the inversion center, but without nodal plane. The same abbreviation applies as in the previous table.}\label{tab:models_no_nodal}
\end{table}
\end{widetext}

\section{DFT calculations for \texorpdfstring{RuO$_2$}{RuO2}}\label{ap:orbital_projection}
\begin{figure}[b]
\begin{center}
\includegraphics[angle=0,width=\linewidth]{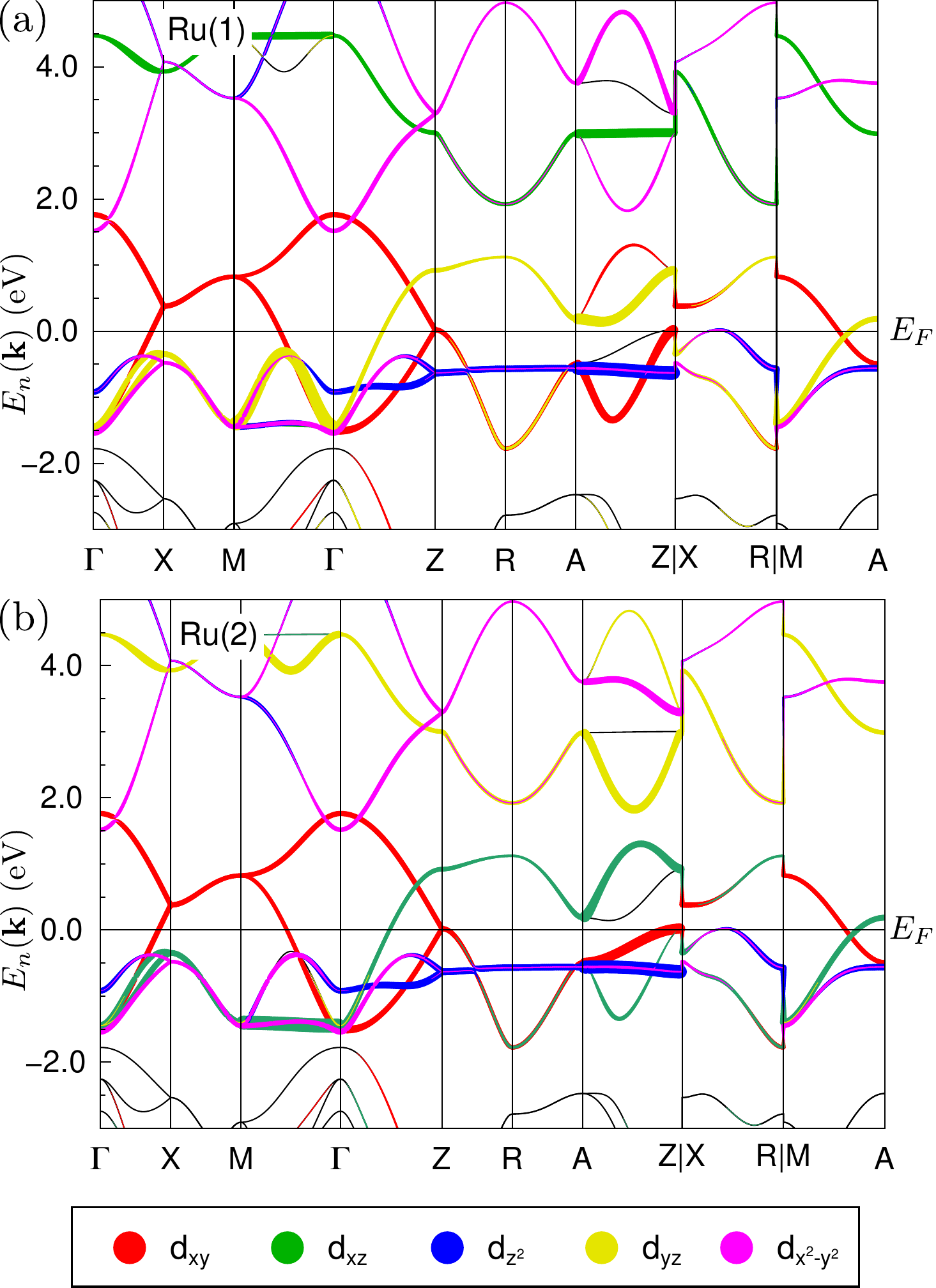}
\caption{DFT band structure (black lines) together with the projection  onto atomic d-orbitals of the Ru(1) atom (a) and the Ru(2) atom (b) with the choice of the local coordinate system as $\hat x=(-1, 1, 0)/\sqrt{2}$ and $\hat z=(1, 1, 0)/\sqrt{2}$ for Ru(1) and $\hat x=(-1, -1, 0)/\sqrt{2}$ and $\hat z=(-1, 1, 0)/\sqrt{2}$ for Ru(2).}
\label{fig:DFTorbitalprojection}
\end{center}
\end{figure}

\emph{Orbital projection of the DFT bands.} In order to construct low-energy minimal models for RuO$_2$, we perform {\it ab initio} calculations using the crystal structure from Ref.~\onlinecite{Project2020}. Within the full-potential local-orbital (FPLO) code, we calculate the electronic structure in the paramagnetic state and examine the orbital character of the band structure by projecting to the d-orbitals in the local octahedral environment of the Ru(1) and Ru(2) atoms, see Fig.~\ref{fig:DFTorbitalprojection}. Focusing on the low-energy dispersion, one can construct minimal one- and two-orbital models that partially describe the respective electronic structure. Constructing a tight-binding model from a Wannier projection allows a Hartree-Fock calculation of the full model finding an altermagnetic instability as discussed in the literature~\cite{Ahn2019}.

\begin{figure}[t]
\begin{center}
\includegraphics[angle=0,width=\linewidth]{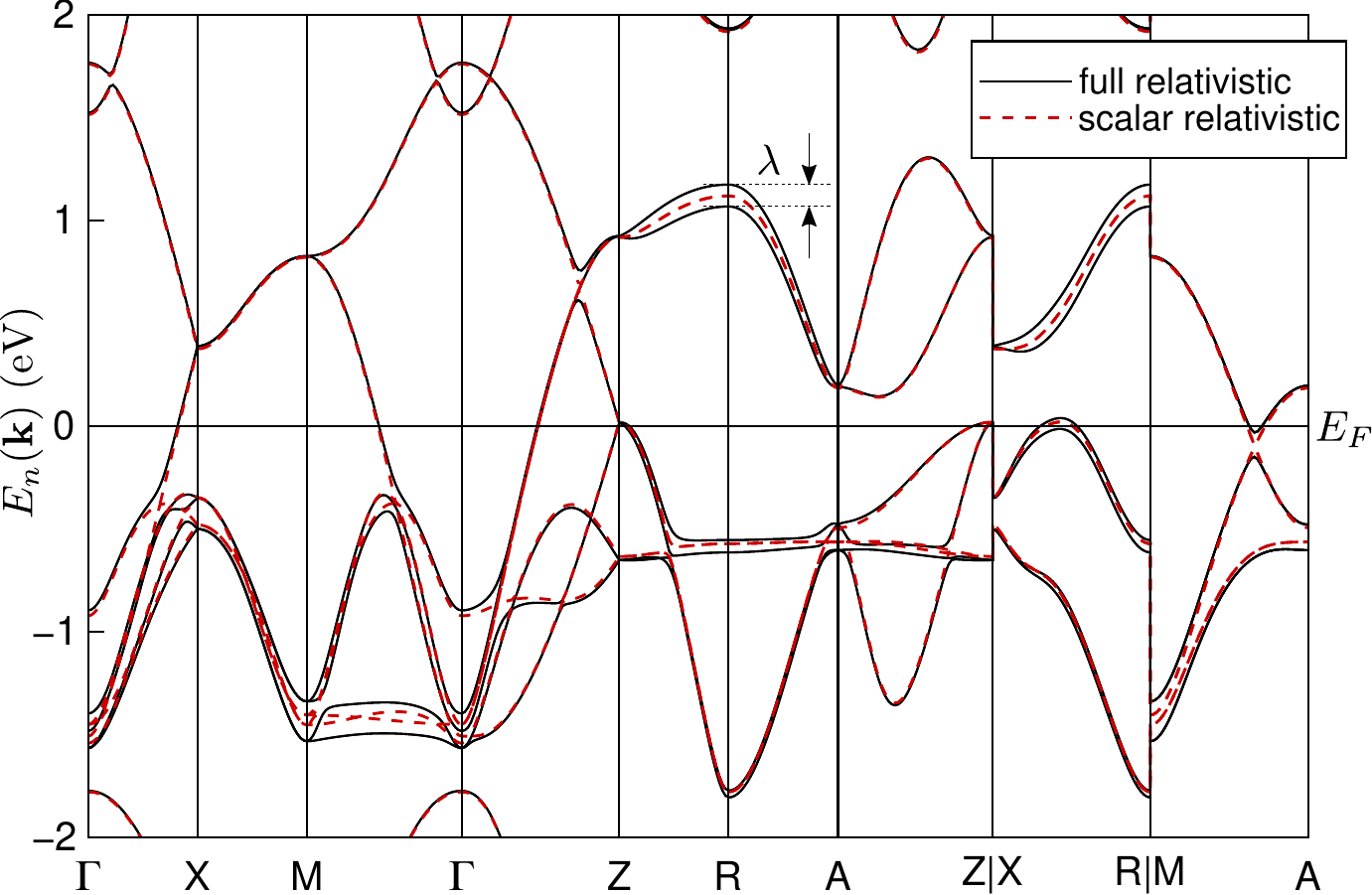}
\caption{Comparison of the band structure of RuO$_2$ from a scalar relativistic DFT calculation (red, dashed) and a fully relativistic calculation (black lines) revealing the splitting of the bands due to SOC.}
\label{fig:relativisticDFT}
\end{center}
\end{figure}

\emph{Relativistic DFT calculations.}
In order to estimate the magnitude of the SOC in RuO$_2$, we perform additional {\it ab initio} calculations in the full relativistic setting of FPLO \cite{Koepernik1999} (version 22.00-62) and plot the band structure for comparison in Fig.~\ref{fig:relativisticDFT}, revealing an effective SOC constant $\lambda=0.1$\;eV, as used in our minimal models of the main text.

\section{The case of \texorpdfstring{MnF$_2$}{MnF2}}\label{ap:DFT_MnF2}
\begin{figure}[h]
\begin{center}
\includegraphics[angle=0,width=\linewidth]{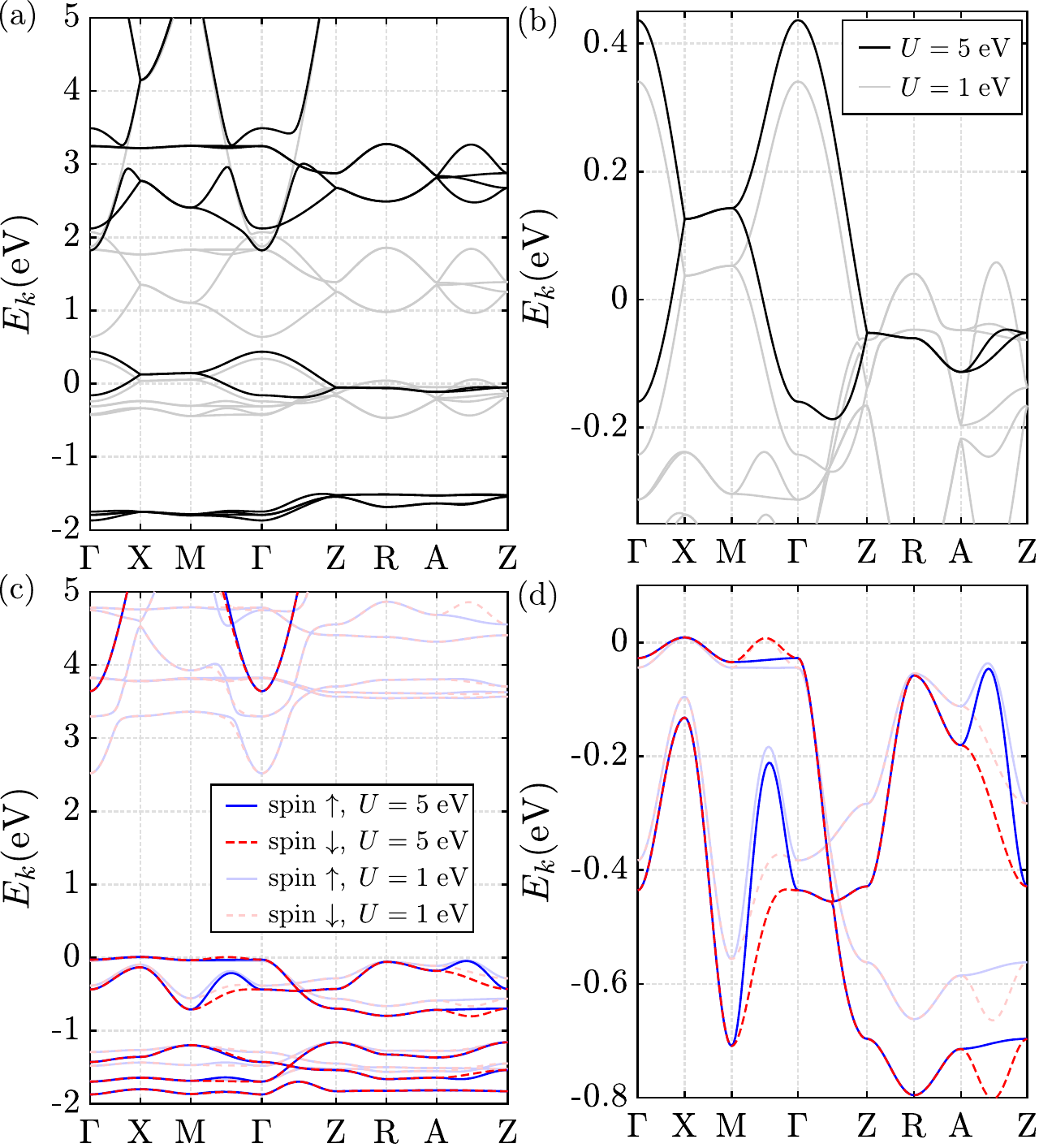}
\caption{DFT calculations for MnF$_2$. (a,b) Electronic structure as obtained from a paramagnetic LDA+U calculation. The bands are shown for two different values of the parameter $U$. (c,d) Electronic structure from a spin-polarized calculation by fixing the total moment to vanish.}
\label{fig:spin_pol_MnF2}
\end{center}
\end{figure}

In Sec.~\ref{sec:minimal_TBH} we have presented the minimal one-orbital model and have demonstrated that it can describe the normal state and altermagnetic band structure of RuO$_2$. Here, we focus on the material candidate MnF$_2$ and compare DFT calculations with the minimal model in Eq.~\eqref{eq:minimal_general_model} relevant for this compound both in the normal state and altermagnetic phase.
In the case of MnF$_2$, DFT calculations have revealed that there is only one orbital close to the Fermi level in the normal state when performing a non spin-polarized calculation~\cite{Yuan2020Jul}. The single-band limit is also shown in Fig.~\ref{fig:spin_pol_MnF2} by introducing an on-site Coulomb interaction $U$.

\begin{figure}[b!]
\begin{center}
\includegraphics[angle=0,width=\linewidth]{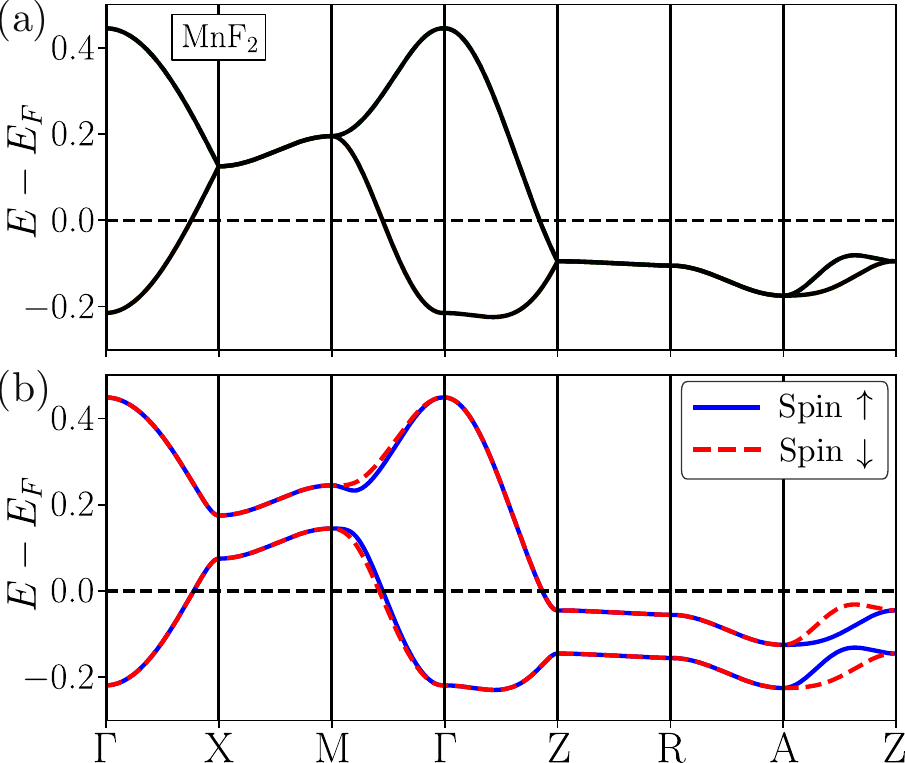}
\caption{Normal state (a) and altermagnetic (b) band structures for MnF$_2$ obtained from the minimal model in Eq.~\eqref{eq:minimal_general_model} with Eqs.~\eqref{eq:xyorb_dispersion}-\eqref{eq:xyorb_hoppings}, hopping parameters detailed in Table~\ref{tab:hoppings_1orb} to reproduce the DFT results (see Fig.~\ref{fig:spin_pol_MnF2}) and $J_z = 0.05$ in (b).
}
\label{fig:MnF2_bands}
\end{center}
\end{figure}

\begin{figure}[tb]
\begin{center}
\includegraphics[width=0.95\linewidth]{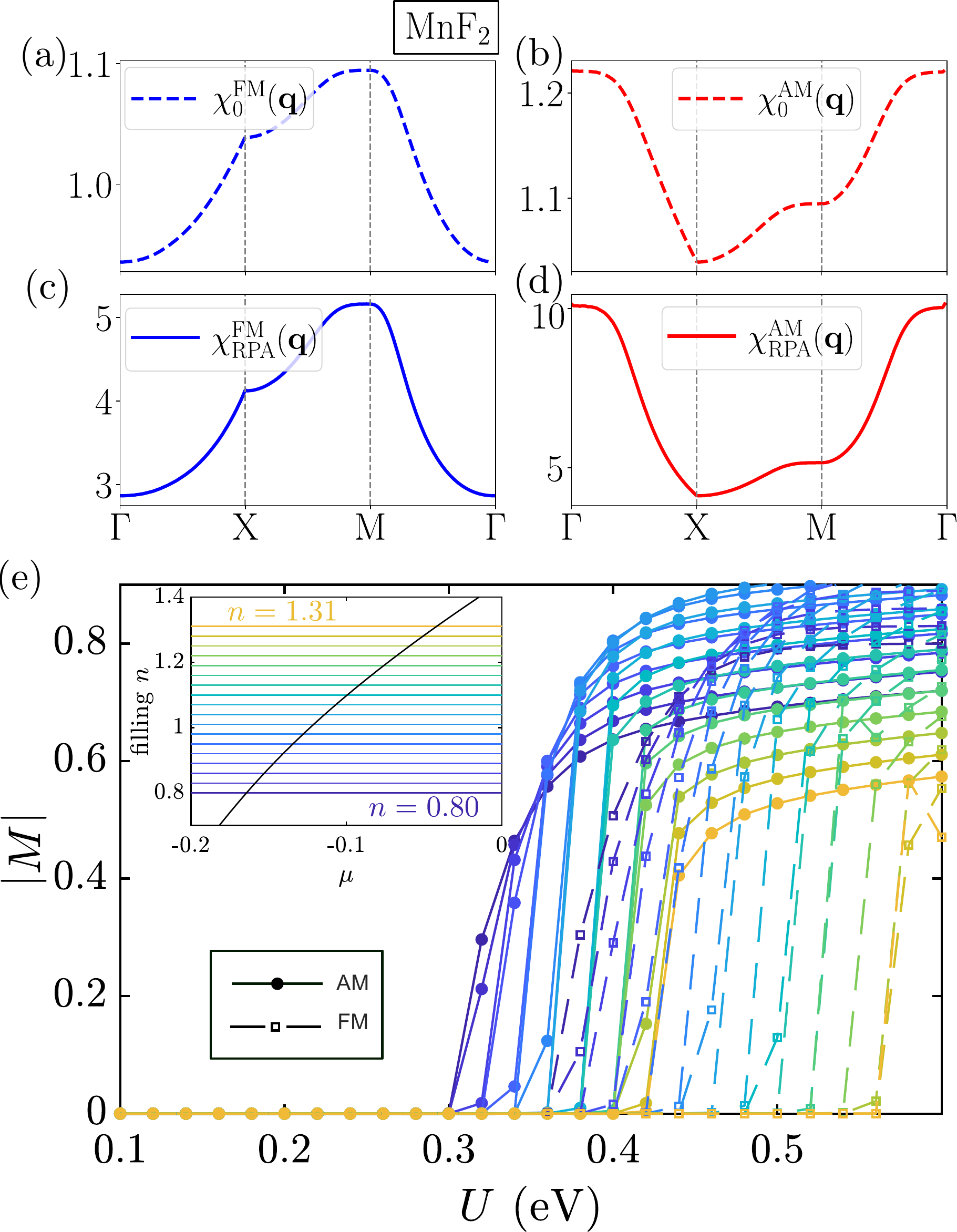}
\caption{(a)-(d) Bare and RPA susceptibilities in the ferromagnetic and altermagnetic channels (see Eqs.~\eqref{eq:RPA_FM},\eqref{eq:RPA_AM}) for MnF$_2$, considering the minimal one-orbital model band structures shown in Fig.~\ref{fig:MnF2_bands}(a), for $U=0.36$, $T=0.02$ and $n_k=60^3$.
(e) Order parameter $|M|=\sum_{\alpha}| n_{\uparrow,\alpha}-n_{\downarrow,\alpha}|$ from a selfconsistent Hartree-Fock calculation at different fillings $n$ where both an altermagnetic and ferromagnetic order parameter can be stabilized, for $T=0.02$ and $n_k=40^3$.}
\label{fig:HF_MnF2}
\end{center}
\end{figure}

\emph{DFT calculations.}
In order to obtain hopping parameters for a minimal model describing the low-energy band structure of MnF$_2$, we adopt the crystal structure from X-ray diffraction in \cite{Baur71} with
space group P4$_2$/mnm (\#136) and lattice constants $a=b=4.873$, $c=3.31$. The Mn atoms are on the 2a Wyckoff position and the F atoms on the 4f position with internal parameter $(0.305,0.305,0)$. We verify the electronic structure as found earlier \cite{Yuan2020Jul} from calculations using the full-potential local-orbital (FPLO) code \cite{Koepernik1999}, version 22.00-62 as well as the full potential linearized augmented plane-wave (LAPW) basis as implemented in Wien2k \cite{Blaha2001}, both using the LDA exchange-correlation functional. Without additional correlations from the LDA+U approach, the entangled Mn d-bands are close to the Fermi level as still visible with weak correlations, see Fig.~\ref{fig:spin_pol_MnF2}(a,b). Once larger correlations are imposed, the d-bands are pushed down and up, leaving a pair of bands from one orbital per Mn close to the Fermi level. Allowing for spin polarization within a calculation in WIEN2k where anti-parallel moments are imposed, we retain two bands for each spin polarization close to the Fermi level, see panel \ref{fig:spin_pol_MnF2}(c,d). A similar state is observed irrespective of whether imposing a finite $U$ or not. However, due to the large local moment, the bands acquire an additional crossing between $\Gamma$ and Z as well as unusual spin shifts along the path between the A and Z points.

\emph{Minimal model.} In Fig.~\ref{fig:MnF2_bands}(a) we display a minimal one-orbital model obtained from Eq.~\eqref{eq:minimal_general_model} relevant for MnF$_2$ in the single-band limit, as shown in Fig.~\ref{fig:spin_pol_MnF2}(a,b) in the normal state DFT calculations. 
Similar to the case of RuO$_2$ in Fig.~\ref{fig:RuO2_bands}, in the normal state the minimal model captures the crossings at the Fermi level and the nodal lines in the band structure.
Note that in the case of MnF$_2$, in the DFT calculations a large $U$ interaction is introduced to obtain the single band crossing the Fermi level. However, allowing for formation of local moments yields a band structure below the Fermi level with different properties (see Ref.~\onlinecite{Yuan2020Jul} and Fig~\ref{fig:spin_pol_MnF2}): the spin splitting in the A-Z line is reversed for one orbital and the spin-up band approaches the spin-down, as opposed to the result obtained from the minimal model, as shown in Fig.~\ref{fig:MnF2_bands}(b).
The more physical picture for MnF$_2$ might be an ordering of large paramagnetic moments occurring at $T=67$ K~\cite{Stout1942Jul} while the compound stays insulating.

\section{Hopping parameters for the minimal models} \label{ap:hoppings}
In order to reproduce the band structures shown in Fig.~\ref{fig:RuO2_bands} for RuO$_2$, Fig.~\ref{fig:bands_FeSb2} for FeSb$_2$, Fig.~\ref{fig:MnF2_bands} for MnF$_2$ and Fig.~\ref{fig:bands_organic} for $\kappa$-Cl, we include in Table~\ref{tab:hoppings_1orb} and Table~\ref{tab:hoppings_2orb} the choice of all hopping parameters. In particular, Table~\ref{tab:hoppings_1orb} details all hoppings for the one-orbital minimal models. Note that the case of $\kappa$-Cl is a two-dimensional model, and therefore some hoppings are not relevant. Table~\ref{tab:hoppings_2orb} includes all hopping parameters for the two-orbital model in the RuO$_2$ case, to obtain the gray bands in Fig.~\ref{fig:RuO2_bands}.

\begin{table}[h]
\caption{Hopping parameters to obtain the band structures for the tetragonal compounds RuO$_2$ in Fig.~\ref{fig:RuO2_bands} and MnF$_2$ in Fig.~\ref{fig:MnF2_bands} (see Eqs.~\eqref{eq:xyorb_dispersion},\eqref{eq:xyorb_hoppings}), and the orthorhombic materials FeSb$_2$ in Fig.~\ref{fig:bands_FeSb2} (Eqs.~\eqref{eq:disperion_FeSb2},\eqref{eq:xyorb_hoppings}) and $\kappa$-Cl in Fig.~\ref{fig:bands_organic} (Eqs.~\eqref{eq:organic_model},\eqref{eq:organic_model1}).}
\label{tab:hoppings_1orb}
\resizebox{\columnwidth}{!}{%
\begin{tabular}{c|ccccccccccc}
\toprule
Tetra. & \multicolumn{2}{c}{$t_1$} & $t_2$ & $t_3$ & \multicolumn{2}{c}{$t_4$} & $t_5$ & $t_6$ & $t_7$ & $t_8$ & $\mu$ \\ \hline 
RuO$_2$ & \multicolumn{2}{c}{-0.05} & 0.7 & 0.5 & \multicolumn{2}{c}{-0.15} & -0.4 & -0.6 & 0.3 & 1.7 & 0.25 \\
MnF$_2$ & \multicolumn{2}{c}{0} & 0.13 & 0 & \multicolumn{2}{c}{-0.02} & 0.015 & 0 & 0.03 & 0.33 & -0.01 \\ \midrule
Ortho. & $t_{1x}$ & $t_{1y}$ & $t_2$ & $t_3$ & $t_{4x}$ & $t_{4y}$ & $t_5$ & $t_6$ & $t_7$ & $t_8$ & $\mu$ \\ \hline
FeSb$_2$ & -0.1 & -0.05 & -0.05 & 0.06 & 0.1 & 0.05 & -0.05 & 0.05 & -0.1 & 0.15 & -0.12 \\
$\kappa$-Cl & 0.08 & -0.01 & - & -0.03 & - & - & - & 0.05 & - & 0.3 & -0.1 \\ \bottomrule
\end{tabular}
}
\end{table}

\begin{table}[h]
\caption{Hopping parameters for the two-orbital model shown in Eqs.~\eqref{eq:xyorb_dispersion}-\eqref{eq:hoppings_2orb} to obtain the band structure in Fig.~\ref{fig:RuO2_bands} (gray line) relevant for RuO$_2$.}
\label{tab:hoppings_2orb}
\begin{tabular}{c|ccccccccc}
\toprule
 & $t_1$ & $t_{2}$ & $t_{3}$ & $t_{4}$ & $t_{5}$ & $t_{6}$ & $t_{7}$ & $t_{8}$ & $\mu$ \\ \hline
RuO$_2$ & 0.18 & -1 & -0.5 & -0.1 & -0.1 & -0.6 & 0 & -0.2 & -3 \\ \midrule
 & $t_9$ & $t_{10}$ & $t_{11}$ & $t_{12}$ & $t_{13}$ & $t_{14}$ & $t_{15}$ & $t_{16}$ & $a_0$ \\ \hline
RuO$_2$ & -0.1 & 0 & -0.2 & 0 & 0 & 0 & 0.1 & 0 & 3 \\ \bottomrule
\end{tabular}
\end{table}

\section{Susceptibility components in sublattice space} \label{ap:suscept_sublatcomp}
\begin{figure}[h]
    \begin{center}
		\includegraphics[angle=0,width=\linewidth]{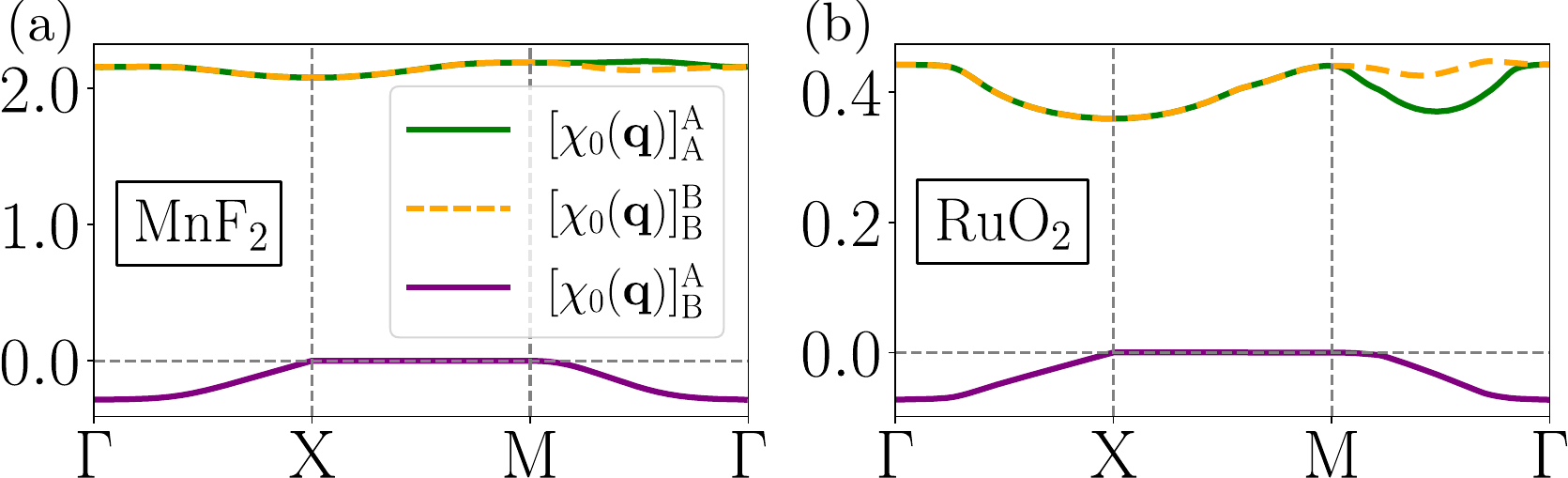}
		\caption{Intra- and inter-sublattice components of the bare susceptibility shown in Eq.~\eqref{eq:bare_suscept_sublat} considering the one-orbital minimal model bands shown in Fig.~\ref{fig:MnF2_bands} for MnF$_2$ (a) and Fig.~\ref{fig:RuO2_bands} for RuO$_2$ (b), with $T=0.02$ and $n_\kv = 60^3$.}
    \label{fig:suscept_sublatticecomp}	
     \end{center}
\end{figure}

The susceptibility in the altermagnetic channel in Eq.~\eqref{eq:RPA_AM} reveals that when the inter-sublattice components become negative the altermagnetic instability is favored. Figure~\ref{fig:suscept_sublatticecomp} shows the sublattice components of the susceptibility for the one-orbital minimal models band structure shown in Fig.~\ref{fig:RuO2_bands}(a) for RuO$_2$ and Fig.~\ref{fig:MnF2_bands}(a) for MnF$_2$, with the bare and RPA susceptibilities shown in Fig.~\ref{fig:HF_RuO2}(a)-(d) and Fig.~\ref{fig:HF_MnF2}(a)-(d), respectively. As seen, for a leading altermagnetic instability the inter-sublattice components become negative along the $\Gamma$-X and M-$\Gamma$ directions, with $[\chi_0 (\qv)]^A_B = [\chi_0 (\qv)]^B_A$. In contrast, when the ferromagnetic instability is leading they are both positive along the same lines. Notably, the susceptibilities $[\chi_0 (\qv)]^A_A$ and $[\chi_0 (\qv)]^B_B$ split in the $\Gamma$-M direction, and the splitting is reversed in the $\Gamma$-M$'$ direction, with  $\textrm{M}=(\pi,\pi,0)$ and $\textrm{M}'=(-\pi,\pi,0)$.

\section{\texorpdfstring{Organic $\kappa$-Cl}{k-Cl}}\label{ap:organic}

The $\kappa$-Cl organic compound belongs to the 2D layer space group L25, and it provides a platform to study 2D altermagnetism for which it is possible to find a weak-coupling mechanism to stabilize the altermagnetic state \cite{Yu2024}. This material differs from the other materials in this work since it lacks inversion symmetry. This lack of inversion symmetry will alter the form of the SOC which is expected to be weak. For these reasons we do not discuss SOC in this section. The anomalous Hall effect and spin current generation has been predicted for this material \cite{Naka2020,Naka2019}. In addition, it has been suggested that the magnetic order can induce finite momentum superconductivity \cite{Sumita2023Nov}.
In these works, the models considered have two dimers in the unit cell, which result in four bands, but due to bonding/antibonding the bands split in two pairs. Therefore, in the large dimerization limit we can focus only on the antibonding set of bands \cite{Kino1996Jul}. We argue that this is sufficient to capture a leading altermagnetic instability.
As a consequence, we can construct the minimal model in Eq.~\eqref{eq:minimal_general_model} by considering Wyckoff positions $(0,0,0)$ and $(1/2,1/2,0)$, which, through arguments similar to FeSb$_2$ gives
\begin{equation}
    \varepsilon_{0,\kv} = t_{1x}  \cos k_x + t_{1y} \cos k_y + t_3 \cos k_x \cos k_y  - \mu,
    \label{eq:organic_model}
\end{equation}
with the hopping terms
\begin{equation}
    \begin{aligned}
    & t_{x,\kv} = t_8 \cos \frac{k_x}{2} \cos\frac{k_y}{2}, \\
    & t_{z,\kv} = t_6 \sin k_x \sin k_y.
    \label{eq:organic_model1}
    \end{aligned}
\end{equation}
Similar to the rutile case in Eq.~\eqref{eq:xyorb_hoppings}, here the $t_{z,\kv} \tau_z$ term also gives rise to the spin splitting with a $d$-wave symmetry. 

\begin{figure}[t!]
\begin{center}
\includegraphics[angle=0,width=.95\linewidth]{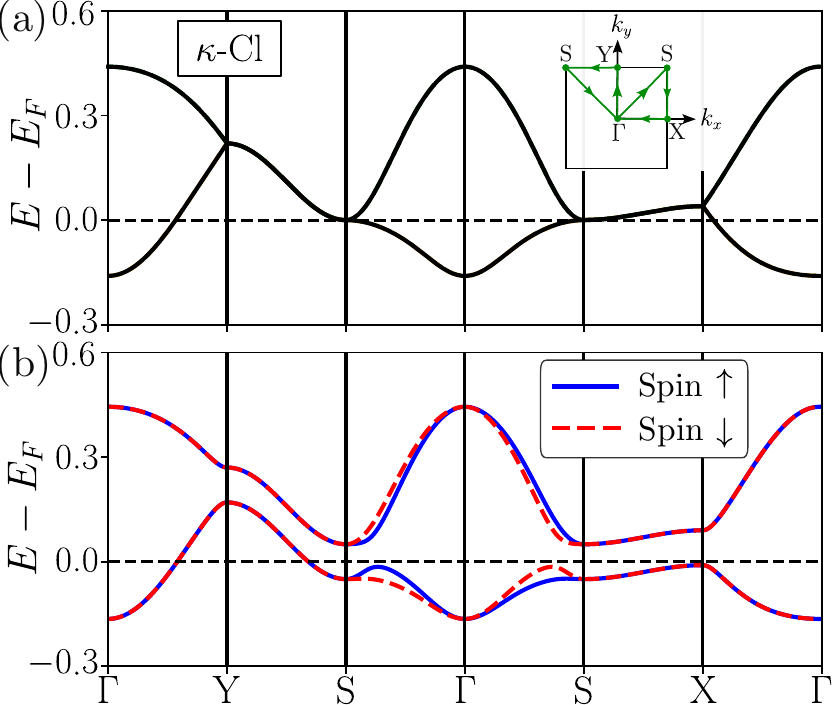}
\caption{Normal state (a) and altermagnetic (b) band structures for $\kappa$-Cl obtained from the minimal model in Eq.~\eqref{eq:minimal_general_model} considering Eqs.~\eqref{eq:organic_model}-\eqref{eq:organic_model1}, with hopping parameters detailed in Appendix~\ref{ap:hoppings} and $J_z = 0.05$ in (b). The BZ path is shown in the inset.
}
\label{fig:bands_organic}
\end{center}
\end{figure}

In Fig.~\ref{fig:bands_organic}(a) we show the band structure by an appropriate choice of the hopping parameters (listed in Appendix~\ref{ap:hoppings}) in Eqs.~\eqref{eq:organic_model}-\eqref{eq:organic_model1}, following the path shown in the inset and assuming that the Fermi level is at the double van Hove singularity at the S-point. The obtained bands are in good agreement with those of Refs.~\onlinecite{Naka2019,Sumita2023Nov}. Figure~\ref{fig:bands_organic}(b) shows the effect of the altermagnetic order parameter in Eq.~\eqref{eq:minimal_general_model}. As seen, again in agreement with previous works \cite{Naka2019,Naka2020,Sumita2023Nov}, the spin splitting is reversed along the $k_x = k_y$ and $k_x = -k_y$ directions, since $\tau_z \sim \sin k_x \sin k_y$.

\begin{figure}[t]
\begin{center}
\includegraphics[width=\linewidth]{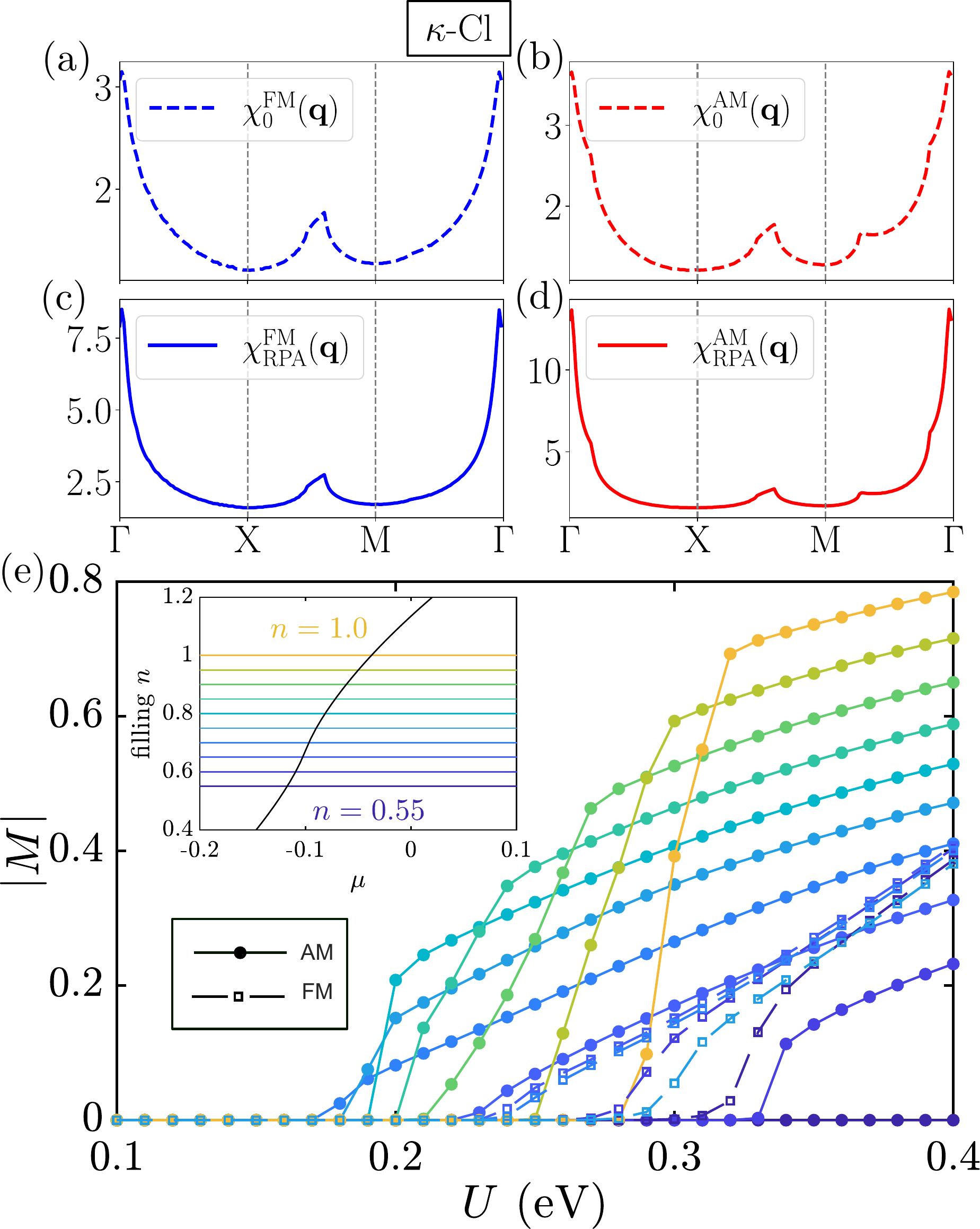}
\caption{(a)-(d) Bare and RPA susceptibilities in the ferromagnetic and altermagnetic channels (see Eqs.~\eqref{eq:RPA_FM},\eqref{eq:RPA_AM}), considering the minimal one-orbital model band structure for $\kappa$-Cl shown in Fig.~\ref{fig:bands_organic}(a), with $U=0.1$ $T=10^{-4}$ and $n_k=400^2$. (e) Order parameter $|M|$ for $\kappa$-Cl calculated within Hartree-Fock using the Hamiltonian in Eq.~(\ref{eq:organic_model1}), with $T=0.002$ and $n_k=300^2$.}
\label{fig:HF_kappaCl}
\end{center}
\end{figure}

As shown in Figs.~\ref{fig:HF_kappaCl}(a)-(d), for this band the van Hove singularity at the Fermi level gives rise to a $\qv \rightarrow 0 $ peak in both the ferromagnetic and altermagnetic bare susceptibilities, but the latter is the leading instability. The Hartree-Fock results in Fig.~\ref{fig:HF_kappaCl}(e) show that the altermagnetic state can be stabilized at a lower value of the Hubbard interaction $U$.  When initializing with ferromagnetic ordering, this state is in principle also stable, and the magnetization sets in gradually with $U$. These results are robust against changes in the overall filling with the modification that the saturation value of the (sublattice) magnetization first decreases and then increases again. This can be understood in terms of a non-monotonic behavior of the density of states (slope of black curve in inset) of Fig.~\ref{fig:HF_kappaCl}.

Examining the model in Eqs.~\eqref{eq:organic_model}-\eqref{eq:organic_model1} shows that a sufficiently large $t_8$ hopping is crucial to obtain a dominant interband susceptibility, in agreement with the analysis of the effect of coincident van Hove singularities in Ref.~\onlinecite{Yu2024}. Motivated by this, in Appendix~\ref{ap:2D_tetragonal_toymodel} we describe a toy model for a 2D tetragonal system inspired by RuO$_2$, giving rise to a leading altermagnetic instability assuming that the Fermi level is also at the coincident van Hove singularities.

\section{Minimal two-dimensional model for altermagnetism in a tetragonal system} \label{ap:2D_tetragonal_toymodel}
In Section~\ref{sec:minimal_TBH} we introduce the general one-orbital model describing altermagnets, and demonstrated in Fig.~\ref{fig:RuO2_bands} that it can reproduce the main features of the bands for RuO$_2$. The dispersion and tight-binding parameters written in Eqs.~\eqref{eq:xyorb_dispersion}-\eqref{eq:xyorb_hoppings} correspond to a 3D system. In this Appendix, we present a minimal 2D model that gives rise to a leading altermagnetic instability, inspired by the RuO$_2$ bands in Fig.~\ref{fig:RuO2_bands}(a) and assuming that the Fermi level is at the coincident van Hove singularities at the M-point~\cite{Yu2024}.

\begin{figure}[t]
    \begin{center}
		\includegraphics[angle=0,width=0.95\linewidth]{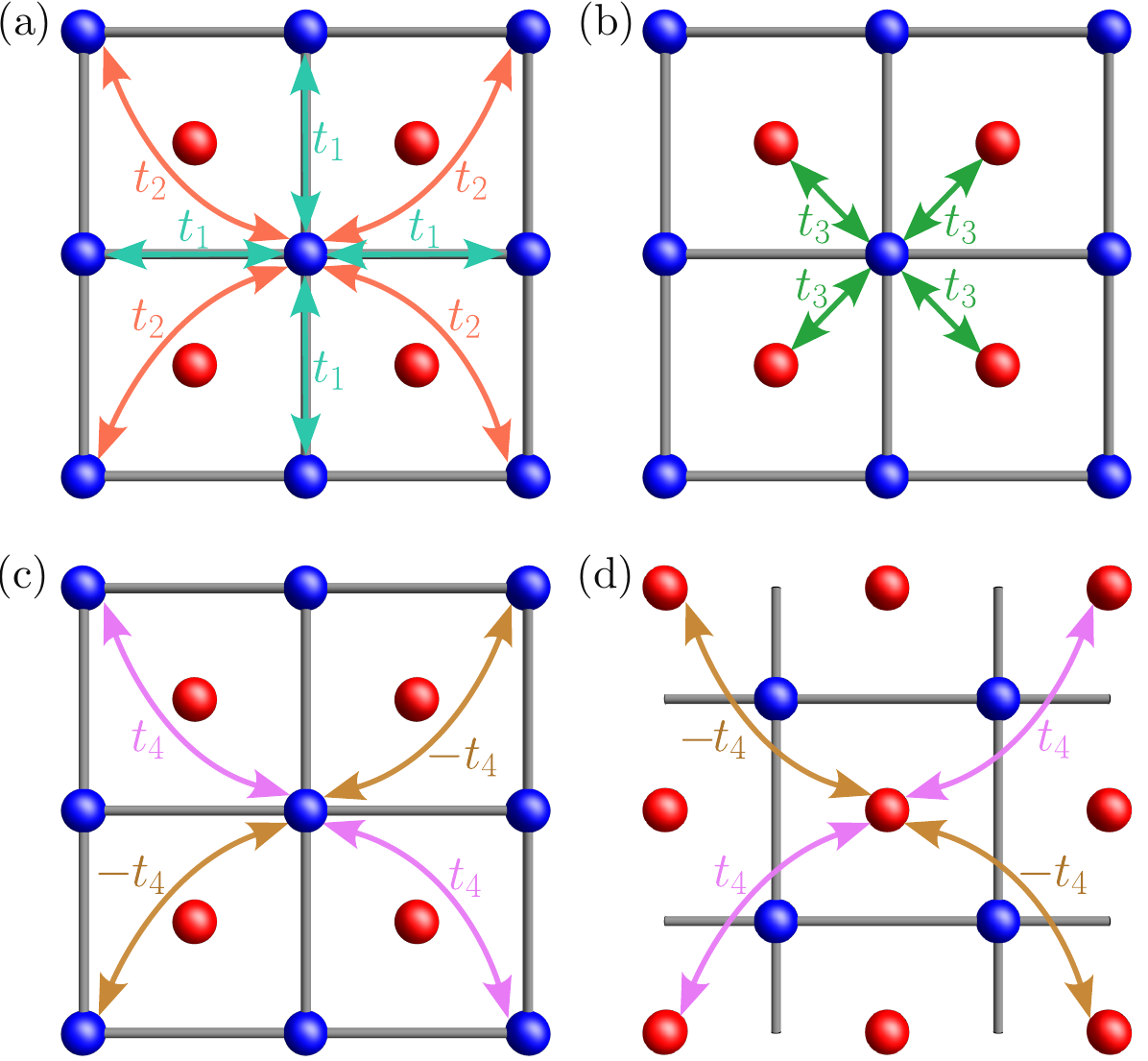}
		\caption{Sketch of the relevant hoppings in the minimal model presented in Eq.~\eqref{eq:2D_minimalmodel_RuO2}, with the red and blue colors representing the two sublattices.}
    \label{fig:Ap_2D_minmodel}	
     \end{center}
\end{figure}

The minimal 2D model for the tetragonal system has the following form:
\begin{equation}
    \begin{aligned}
    H_{2D} & = t_1 (\cos k_x + \cos k_y) + t_2 \cos k_x \cos k_y - \mu \\
    &+ t_3 \cos \frac{k_x}{2}\cos \frac{k_y}{2} \tau_x + t_4 \sin k_x \sin k_y \tau_z,
    \end{aligned}
    \label{eq:2D_minimalmodel_RuO2}
\end{equation}
where we have omitted the SOC terms for simplicity. This model is obtained by evaluating
the 3D model in Eqs.~\eqref{eq:xyorb_SOC}, \eqref{eq:xyorb_dispersion} and \eqref{eq:xyorb_hoppings} at $k_z=0$. The hoppings in Eq.~\eqref{eq:2D_minimalmodel_RuO2} are illustrated in Fig.~\ref{fig:Ap_2D_minmodel}.

\begin{figure}[h]
    \begin{center}
		\includegraphics[angle=0,width=0.95\linewidth]{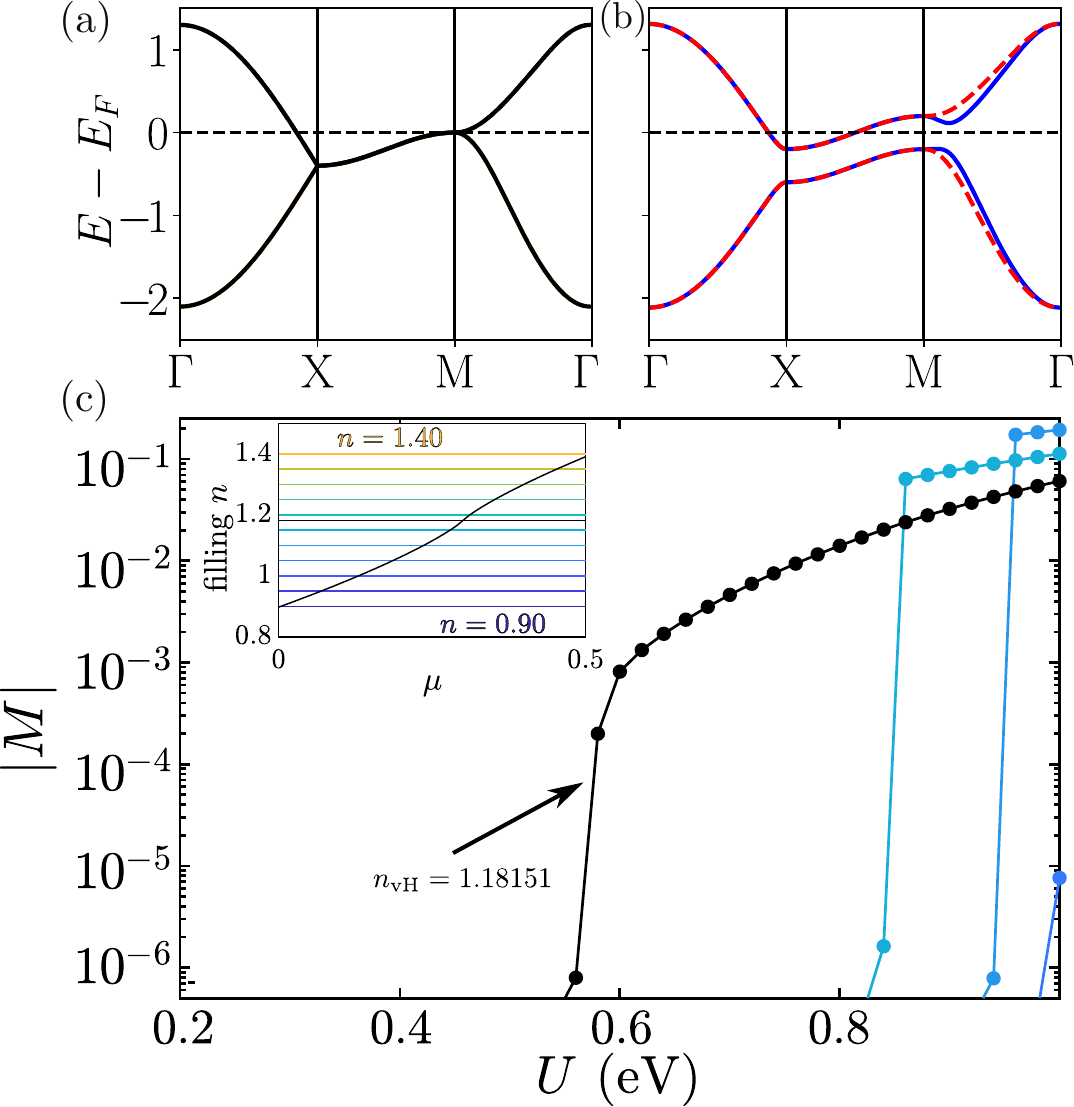}
		\caption{Normal state (a) and altermagnetic band structure (b) obtained from the 2D toy model in Eq.~\eqref{eq:2D_minimalmodel_RuO2} inspired by the RuO$_2$ bands in Fig.~\ref{fig:RuO2_bands}. The hopping parameters are $\{t_1,t_2,t_3,t_4,\mu\} = \{-0.1, 0.1,1.7,0.3,0.3\}$, with $J_z = 0.2$ for the magnetic order parameter in (b) [see Eq.~\eqref{eq:minimal_general_model}]. (c) Order parameter $|M|$ for the bands in (a) calculated within Hartree-Fock using the Hamiltonian in Eq.~(\ref{eq:2D_minimalmodel_RuO2}), with $T=0.0001$ and $n_k=1200^2$.}
    \label{fig:RuO2bands_2D_minmodel}	
     \end{center}
\end{figure}

In Fig.~\ref{fig:RuO2bands_2D_minmodel} we show the band structure obtained from the minimal model in Eq.~\eqref{eq:2D_minimalmodel_RuO2}, both in the normal state and the altermagnetic state. We have verified with Hartree-Fock calculations that this model gives rise to a leading altermagnetic instability, as shown in Fig.~\ref{fig:RuO2bands_2D_minmodel}(c). Importantly, the altermagnetic instability is stabilized at a lower value of the interaction $U$ compared to the ferromagnetic ordering. In addition, Fig.~\ref{fig:RuO2bands_2D_minmodel}(c) shows two more details of this two dimensional toy model. First, for variations slightly away from the van Hove filling the critical $U$ for the altermagnetic ordering becomes larger. Second, the order parameter has a steep onset (note the logarithmic scale) and quickly reaches values where the electronic structure beomes insulating as also found in a similar model in Ref.~\onlinecite{Maier2023}.

\section{Topological surface state with altermagnetic order parameter \texorpdfstring{$\tau_z \sigma_z$}{tauz sigmaz}} \label{ap:topological_states}

As an example of the utility of our minimal models, we develop conditions under which altermagnetism can lead to topologically protected edge states. As discussed in the main text, the altermagnetic order parameter $\tau_z\sigma_z$ preserves the mirrors $M_x, M_y, M_z$. Consequently, there is a vanishing anomalous Hall effect and Berry curvature. However, on the high-symmetry plane, it can exhibit non-trivial topological properties akin to those of  topological crystalline insulators~\cite{fu:2011}.

For experiments preserving $M_z$, it is worth studying the high-symmetry plane $k_z=0$ and $k_z=\pi$. We take $k_z=0$ plane as an example. The minimal Hamiltonian for RuO$_2$ becomes ($d_{xy}$ orbitals only):
\begin{equation}
	\begin{split}
	H(k_z=0)&= t_8 \cos \frac{k_x}{2}\cos\frac{k_y}{2}\tau_x\\
	&\phantom{=}+(t_6+t_7) \sin k_x \sin k_y\tau_z\\
	&\phantom{=}+\lambda_z \cos \frac{k_x}{2} \cos \frac{k_y}{2} (\cos k_x - \cos k_y)\tau_y\sigma_z\\
	&\phantom{=}+J_1\tau_z\sigma_z+J_2\sin\frac{k_x}{2}\sin\frac{k_y}{2}\tau_x\sigma_z\\
	&\phantom{=}+J_3\sin\frac{k_x}{2}\sin\frac{k_y}{2}(\cos k_x-\cos k_y)\tau_y.
	\end{split}
\end{equation}

\begin{figure}[t]
    \begin{center}
		\includegraphics[width=7cm]{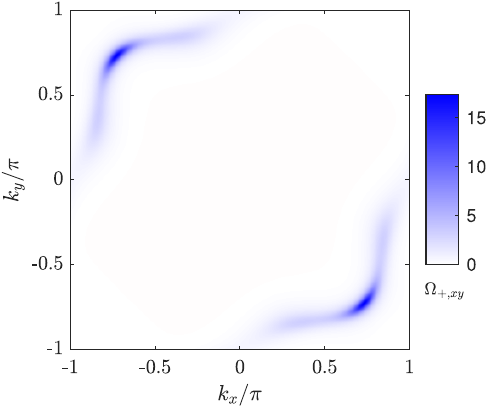}
		\caption{Berry curvature in the $\sigma_z=+$ sector on the $k_z=0$ plane. We take $t_8=1$, $t_6+t_7=1/2$, $\lambda_z=1/4$, and $J_1=J_2=J_3=1/4$. Here the integral of the Berry curvature over the $k_z=0$ plane is $4\pi$.}
    \label{fig:Berry_sector}	
     \end{center}
\end{figure}

Here, the $J_2$ and $J_3$ terms have the same symmetry as the order parameter $J_1\tau_z\sigma_z$, and they have a $\cos\frac{k_z}{2}$ factor in 3D. Since momenta are invariant under the $M_z\propto \sigma_z$ symmetry, the Hamiltonian can be block-diagonalized based on the eigenvalues of $M_z$. These blocks correspond to the $\sigma_z=\pm$ sectors. Since these sectors are independent, we can define the Berry curvature in each sector separately. As the mirror operators $M_x$ and $M_y$ anti-commute with $M_z$, they interchange the two sectors. While analyzing a given sector, these two mirror symmetries are absent. Therefore, the Berry curvature on the Fermi surface in each sector is generically non-zero, as shown in Fig.~\ref{fig:Berry_sector}. To ensure that the total Berry curvature vanishes, these two Berry curvatures need to be opposite. This implies that, on the $x=0$ plane, spin-up and spin-down electrons with $k_z=0$ generically propagate along opposite $y$-directions.

We now illustrate the possibility of chiral surface bound states. Focusing on the $\sigma_z=+$ sector, with generic parameter $J_1$, the resulting two-band model is fully gapped. As the parameter $J_1$ is tuned, point-wise gap closings can happen in $(J_1,k_x,k_y)$ space. The closings are at $J_1=0$ and $J_1=4(t_6+t_7)\frac{t_8/J_2}{(1+|t_8/J_2|)^2}$. Berry phase is thus non-zero for $J_1$ between the two closings, which implies the existence of chiral surface states. On the $x=0$ surface, these chiral surface modes have $k_z=0$, and propagate along the y-direction. In the $\sigma_z=-$ sector, the Berry phase is opposite, and the surface bound states propagate in the opposite direction. These pairs of chiral states resemble the quantum spin Hall effect, but are governed by mirror symmetries and located at high-symmetry planes.

\section{RPA and selfconsistent Hartree-Fock results for \texorpdfstring{FeSb$_2$}{FeSb2}} \label{ap:HF_organic_FeSb2}
\begin{figure}[b]
\begin{center}
\includegraphics[width=\linewidth]{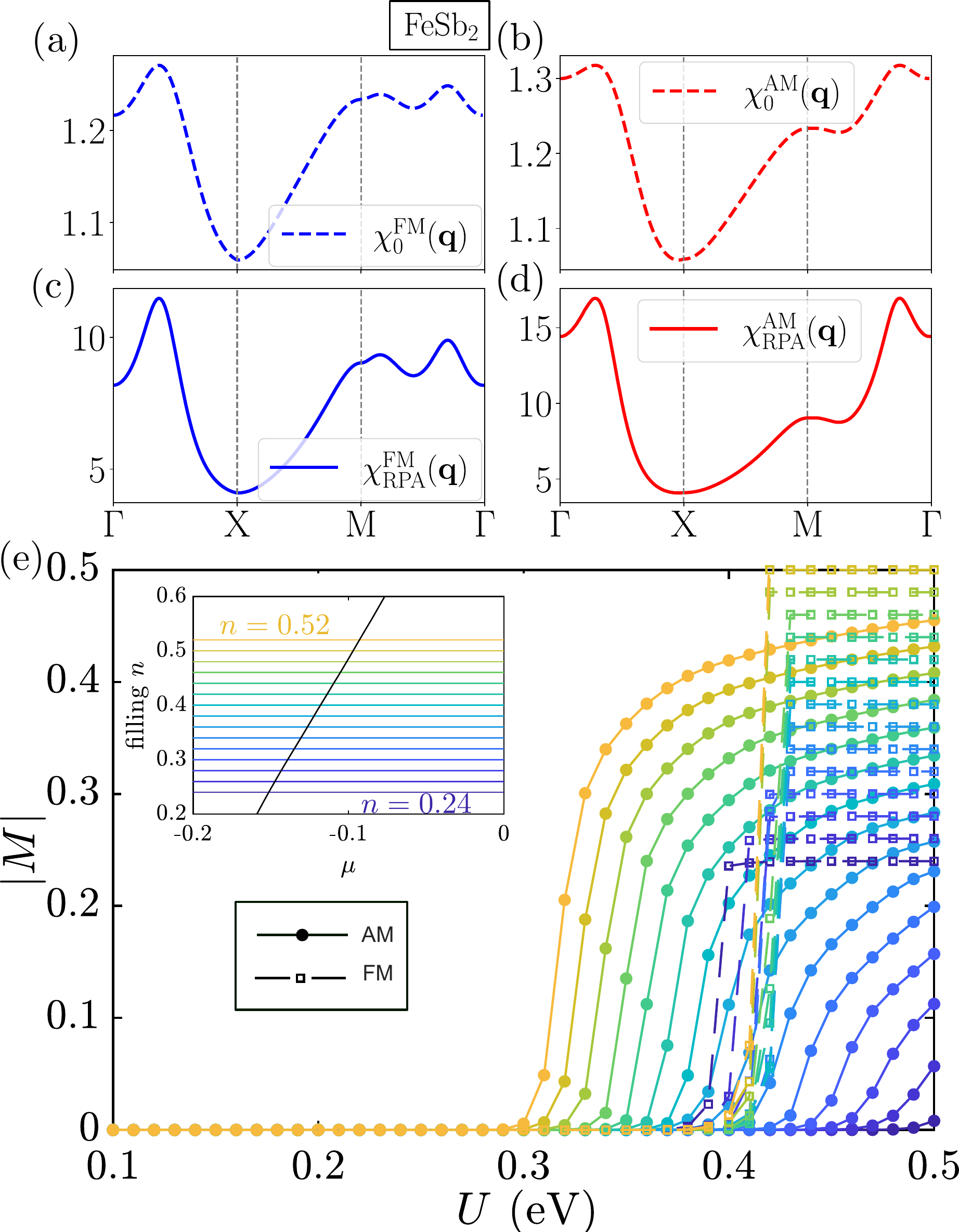}
\caption{(a)-(d) Bare and RPA susceptibilities in the ferromagnetic and altermagnetic channels (see Eqs.~\eqref{eq:RPA_FM},\eqref{eq:RPA_AM}), considering the minimal one-orbital model band structure inspired on FeSb$_2$ shown in Fig.~\ref{fig:bands_FeSb2}(a), with $U=0.35$, $T=0.02$ and $n_k=60^3$. (e) Order parameter for FeSb$_2$ obtained within Hartree-Fock from the Hamiltonian in Eq.~(\ref{eq:disperion_FeSb2}), with $T=0.002$ and $n_k=40^3$.}
\label{fig:HF_FeSb2}
\end{center}
\end{figure}

In this Appendix, we present results from RPA analysis and selfconsistent Hartree-Fock calculations using the minimal model describing the low-energy electronic structure of {FeSb$_2$} where spin-resolved densities are allowed as order parameters.
Figure~\ref{fig:HF_FeSb2}(a)-(d) display the bare and RPA susceptibilities in the ferromagnetic and altermagnetic channels using the one-orbital minimal model in Eq.~\eqref{eq:minimal_general_model} for FeSb$_2$, with the bands shown in Fig.~\ref{fig:bands_FeSb2}(a). As discussed in Sec.~\ref{sec:TBM_AMcandidates}A, the RPA analysis shows that the altermagnetic susceptibility is the leading instability, but diverges at a finite $\qv$. 

As seen from the Hartree-Fock results in Fig.~\ref{fig:HF_FeSb2}(e), the altermagnetic state can be stabilized at a lower value of the Hubbard interaction $U$. When initializing with ferromagnetic ordering, this state is in principle also stable, and it yields a fully polarized ferromagnet beyond its critical $U$. These results are robust against changes in the overall filling with the modification that the saturation value of the (sublattice) magnetization increases with filling and the critical $U$ for FeSb$_2$ decreases (when approaching half filling).

\bibliography{AM}

\end{document}